\documentclass[twocolumn,superscriptaddress,amsfont,amssymb,amsmath, showpacs,balancelastpage]{revtex4-1}
\usepackage{CJKutf8}

\usepackage{amsthm}
\newtheorem{theorem}{Theorem}
\newtheorem{corollary}{Corollary}[theorem]
\newtheorem{lemma}[theorem]{Lemma}

\usepackage[utf8]{inputenc}
\usepackage{amsmath,amssymb,amsfonts,stmaryrd}
\usepackage{physics}
\usepackage{dsfont}
\usepackage{graphicx}
\usepackage{xcolor}
\usepackage{graphicx}
\usepackage{cancel}
\usepackage{natbib}
\usepackage{color}
\usepackage{tikz}
\usepackage{mathrsfs}
\usepackage{relsize}
\usepackage{multirow}
\usepackage[linktocpage]{hyperref}
\usepackage[all]{xy}
\usepackage[export]{adjustbox}
\hypersetup{colorlinks=true,citecolor=blue,linkcolor=blue, urlcolor=blue, breaklinks=true}

\usepackage{float}

\usepackage{bm} %allows for bold math type
\usepackage{subfigure} %allows for 2x2 figures
\usepackage{environ}
\usepackage{url}
\usepackage[margin=0.75in]{geometry}
\usepackage{ulem}

\usepackage{mathtools}

\NewEnviron{eqs}{%
\begin{equation}\begin{split}
    \BODY
\end{split}\end{equation}}

\definecolor{dkgreen}{rgb}{0,0.5,0}
\usetikzlibrary{calc,arrows.meta}

\definecolor{xcol}{RGB}{200,40,40}
\definecolor{zcol}{RGB}{30,100,190}
\definecolor{cupA}{RGB}{230,140,20}
\definecolor{cupB}{RGB}{60,160,80}
\definecolor{cupC}{RGB}{140,80,190}

\theoremstyle{definition}
\newtheorem{definition}{Definition}

\theoremstyle{remark}

\date{\today}

\begin{document}

\begin{CJK*}{UTF8}{bsmi}

\title{
Toward Universal Fault-Tolerant Quantum Computation \\
with the Self-Correcting 6D 2-Form $\mathbb{CP}^3$ Code
}

\author{Po-Shen Hsin}
\affiliation{Department of Mathematics, King’s College London, Strand, London WC2R 2LS, UK}
\email[E-mail: ]{po-shen.hsin@kcl.ac.uk}

\author{Ryohei Kobayashi}
\affiliation{Department of Applied Physics, University of Tokyo, Bunkyo-ku, Tokyo 113-8656, Japan}
\email[E-mail: ]{ryohei.k@ap.t.u-tokyo.ac.jp}

\author{Zhenghan Wang}
\affiliation{Department of Mathematics, University of California, Santa Barbara, California 93106, USA}
\email[E-mail: ]{zhenghwa@math.ucsb.edu}

\date{\today}
\begin{abstract}
We study the 2-form $\mathbb{CP}^3$ code, a 6D loop homological code that encodes exactly one logical qubit on a space of $\mathbb{CP}^3$ topology. We prove that it is a self-correcting quantum memory on families of refined lattices: below a temperature independent of the lattice size, the logical state is preserved under weak thermal coupling for an exponentially long time in the linear size of the lattice. The code also has a logical CCZ gate implemented by a constant-depth local unitary built from the triple cup product. The 4D 2-form $\mathbb{CP}^2$ code admits all Clifford gates as constant-depth local unitaries, and we propose universal quantum computation by code switching between the two codes, showing in the absence of noise that the switch preserves the CCZ logical state up to a Pauli frame. The smallest known triangulation of $\mathbb{CP}^3$ gives an explicit $[[783,1,4]]$ code, on which the CCZ circuit has depth 27. In circuit-level simulations of this small code, a single-block proxy of magic-state preparation with correction-and-detection decoding has an accepted error consistent with cubic rather than quadratic suppression over the sampled range.
\end{abstract}

\maketitle

\end{CJK*}

\bigskip\bigskip
\begin{abstract}
...
\medskip
\noindent
\end{abstract}

\bigskip \bigskip \bigskip
 
% \date{\today}

\bigskip

% \eject

\tableofcontents

\unitlength = .8mm

\setcounter{tocdepth}{3}

\bigskip

\section{Introduction}
\label{sec:intro}

Fault-tolerant quantum computation requires both a reliable quantum memory and a universal set of logical operations, and achieving both with a low overhead in qubits and syndrome measurements remains a central challenge. Quantum error-correcting codes make this possible by discretizing errors so that they can be detected and corrected (see e.g.~\cite{gottesmanbook}), and error correction has now been demonstrated on several hardware platforms (see e.g.~\cite{Bluvstein:2023logical,google25,Paetznick:2026cad}). All roadmaps to fault-tolerant universal quantum computers rely on such codes, so the search for robust, high-performing codes with good logical gates is central to the construction of a utility-scale quantum computer.  

Topological codes such as the surface code have been at the forefront of hardware implementations. The surface code, however, lacks two desirable properties: it is neither single-shot~\cite{Bombin:2014ksv}, i.e.\ correctable from a single round of noisy syndrome measurement, nor self-correcting, i.e.\ protected by an energy barrier that grows with system size.  Over the last few years, spectacular progress in neutral-atom and trapped-ion platforms has liberated codes from planar constraints and opened the door to exploring high-dimensional, high-performing ones.  Numerical simulations show that the 4D single-shot loop-only toric code is competitive~\cite{ms25}.  Moreover, 6D topological codes can be made into a self-correcting quantum computer with transversal gates and measurements~\cite{Bombin_2013}.  For topological quantum codes, the Bravyi-K\"onig theorem revealed a deep relation between the dimension of spaces and the hierarchy of logical gates~\cite{BK}.  Aspects of this connection manifest themselves in the toric codes of dimensions 2, 4, and 6 on spaces $T^2, T^4$, and $T^6$: in 2D, the toric code is neither single-shot nor self-correcting; in 4D, the loop-only toric code is single-shot and self-correcting; but only in 6D and higher can a self-correcting quantum computer be constructed from logical gates implemented by constant-depth local unitaries and fault-tolerant single-qubit measurements. 

Quantum error-correcting codes in high spatial dimensions offer two
resources that two- and three-dimensional codes lack simultaneously:
\textit{self-correction}, enabled by extended excitations that are suppressed at finite temperature, and \textit{non-Clifford gates implemented by constant-depth local
unitaries}, which arise from triple cup products and are
obstructed in low dimensions~\cite{BK}. Loop toric codes on $2N$-dimensional tori realize both, but they
encode multiple logical qubits, which makes logical synthesis costly and
complicated, and they admit only a partial set of Clifford gates implemented by
constant-depth local unitaries. It is therefore of both theoretical interest and practical value
to study homological codes on other manifolds: to find
single-logical-qubit analogues, to identify topologies that support
larger sets of such Clifford gates, and to understand how the
underlying manifold influences their quantum error-correction
properties. For 4D, codes from cellulations of $\mathbb{CP}^2$ are the subject of
related ongoing work~\cite{BW}.

In this paper we develop a 6D counterpart of the self-correcting
quantum computer of~\cite{Bombin_2013} with a single logical qubit on a
closed manifold. We study the loop homological code on the complex
projective space $\mathbb{CP}^N$, which we call the 2-form
$\mathbb{CP}^N$ code, and develop the case $N=3$ from topological field
theory down to explicit code constructions with syndrome-extraction
circuits and decoders. The choice of $\mathbb{CP}^N$ is natural on three counts:
its homology supplies exactly one logical qubit; its K\"ahler form
generates a cohomology ring with $\int_{\mathbb{CP}^N}\omega^N=1$, which
leads to diagonal logical gates generated by cup-product circuits;
and for $N\ge2$ all its excitations are extended, as
self-correction and single-shot error correction require.

Our main results are as follows.
\begin{enumerate}
\item \textit{Self-correction} (Theorem~\ref{thm:selfcorrection} in Sec.~\ref{sec:selfcorrection}). 
For $N\ge2$, we prove self-correction for the 2-form $\mathbb{CP}^N$ code on the family of edgewise subdivisions $K_m$ of any fixed triangulation. At a fixed temperature below a size-independent threshold, the theorem gives a memory-time lower bound of $e^{\Omega(m)}$ under weak thermal coupling, exponential in the lattice’s linear size.
The initial state is a specified thermal encoding: the desired code state dressed by Pauli errors sampled from the Gibbs distribution conditioned on the decoder’s trivial logical sector. This initialization is an assumption of the theorem; the same bound is not established for an exact code state. The proof in Appendix~\ref{app:selfcorrection} adapts the argument of Ref.~\cite{doi:10.1142/S1230161210000023} for the 4D toric code.

\item \textit{A non-Clifford gate} (Sec.~\ref{sec:ccz}). The 6D 2-form $\mathbb{CP}^3$ code has a logical $CCZ$ gate between three copies, implemented by a constant-depth local unitary of physical $CCZ$ gates built from the triple cup product. 
\item \textit{All Clifford gates} (Secs.~\ref{sec:czgate}--\ref{sec:hgate} and Table~\ref{tab:gates}). The 4D 2-form $\mathbb{CP}^2$ code has $S$, $HSH$ and $CZ$ gates implemented by constant-depth local unitaries, and hence $H=S\cdot HSH\cdot S$, so that all Clifford gates are available in constant depth. On a triangulation, the $HSH$ gate uses a local cochain model of the dual lattice that we construct.
\item \textit{Code switching} (Sec.~\ref{sec:universal}). Switching from the 6D code to the 4D code, by measuring Pauli $X$ on the 2-cells outside a $\mathbb{CP}^2$ subcomplex, turns the output of the $CCZ$ gate into $|CCZ\rangle$ magic states for the 4D code. In the absence of noise, this switch preserves the logical state up to a Pauli frame computed from the outcomes. Because both codes have the metacheck structure that motivates single-shot recovery (Sec.~\ref{sec:singleshot}), the resulting protocol for universal computation is expected to need only $O(1)$ rounds of error correction per logical operation (Sec.~\ref{sec:universal} and Fig.~\ref{fig:protocol}).
\end{enumerate}

We then make the 6D code concrete. The smallest known triangulation of $\mathbb{CP}^3$~\cite{Bagchi:2012cp3} gives a $[[783,1,4]]$ code whose parameters we compute exactly (Theorem~\ref{thm:distance18}), and systematic subdivisions raise the distance at larger size (Sec.~\ref{sec:lattices}). 
The syndrome-extraction circuit measures the $Z$-checks followed by the $X$-checks, using one bare ancilla per check.  We benchmark $\bar X$ and $\bar Z$ separately by preparing the corresponding logical basis state, applying one noisy syndrome-extraction round, and measuring every data qubit (Sec.~\ref{sec:simulation}). The noise model includes depolarizing noise on every CNOT and bit flips on every measurement outcome.
The $\bar Z$ benchmark benefits from the large $X$ distance, $d_X\ge91$: its logical error rate drops sharply as the physical error rate falls below a few percent. The $\bar X$ benchmark is limited by the weight-$4$ $Z$-type logical operators. Its logical error rate scales as $p^2$, but the prefactor remains large on this distance-$4$ lattice.

Finally, we apply correction-and-detection decoding~\cite{Paetznick:2026cad} to the preparation of $|CCZ\rangle$ magic states. Single-block proxy simulations on the $[[783,1,4]]$ code show that postselection improves the error rate among accepted attempts, at the cost of a lower acceptance probability (Sec.~\ref{sec:cczprep}).
These results establish the small code as a proof of concept for the circuits and decoders. Its weight-$4$ $Z$-type logical operators limit both $\bar X$ protection and magic-state preparation, making larger $d_Z$ the main practical goal. Thresholds and the performance of larger lattices remain to be studied. Such studies will help assess the prospects for fault-tolerant computation with these codes on platforms with flexible connectivity, such as neutral atoms and trapped ions (Sec.~\ref{sec:hardware}).

The paper is organized as follows. Section~\ref{sec:CPN} defines the 2-form $\mathbb{CP}^N$
code. Section~\ref{sec:memory} states the self-correction theorem and describes single-shot error
correction via metachecks. Section~\ref{sec:gates} constructs the
Clifford gates in the 2-form $\mathbb{CP}^2$ code and the CCZ gate in
the 2-form $\mathbb{CP}^3$ code, implemented by constant-depth local unitaries, and
assembles a protocol for universal computation via code switching. Section~\ref{sec:practical} presents the
codes on explicit lattices, circuit-level
simulations, and the correction-and-detection magic-state preparation. Section~\ref{sec:colorcode} compares the scheme with the 6D
color code of Ref.~\cite{Bombin_2013}, and Section~\ref{sec:discussion}
discusses open problems. 
Appendix~\ref{app:memory} describes the detailed proof for the self-correction of the $\mathbb{CP}^3$ code.
Appendix~\ref{app:sim} explains step by step how
the logical error rates are obtained.

\section{2-Form $\mathbb{CP}^N$ Code: Loop Homological Code on $\mathbb{CP}^N$ for $N\geq 2$}
\label{sec:CPN}
\begin{figure*}[t]
\centering
\begin{tikzpicture}[x={(1cm,0cm)},y={(0.5cm,0.36cm)},z={(0cm,1cm)},scale=1.3,line join=round]
\begin{scope}
\fill[xcol!22,draw=gray!70] (0,0,0)--(0,1,0)--(0,1,1)--(0,0,1)--cycle;
\fill[xcol!22,draw=gray!70] (0,0,0)--(-1,0,0)--(-1,0,1)--(0,0,1)--cycle;
\fill[xcol!22,draw=gray!70] (0,0,0)--(1,0,0)--(1,0,1)--(0,0,1)--cycle;
\fill[xcol!22,draw=gray!70,fill opacity=0.85] (0,0,0)--(0,-1,0)--(0,-1,1)--(0,0,1)--cycle;
\draw[line width=1.6pt] (0,0,0)--(0,0,1);
\node[right] at (0.02,0,0.5) {$e$};
\node at (0.55,0,0.62) {\small$X$};
\node at (-0.55,0,0.62) {\small$X$};
\node at (0,0.62,0.85) {\small$X$};
\node at (0,-0.55,0.32) {\small$X$};
\node at (0,0,-0.55) {(a) $X$-check ${\cal S}^X_e$};
\end{scope}
\begin{scope}[xshift=2.7cm]
\fill[zcol!14,draw=gray!60,dashed] (0,1,0)--(1,1,0)--(1,1,1)--(0,1,1)--cycle;
\fill[zcol!14,draw=gray!60,dashed] (0,0,0)--(0,1,0)--(0,1,1)--(0,0,1)--cycle;
\fill[zcol!14,draw=gray!60,dashed] (0,0,0)--(1,0,0)--(1,1,0)--(0,1,0)--cycle;
\fill[zcol!26,draw=gray!80,fill opacity=0.9] (0,0,0)--(1,0,0)--(1,0,1)--(0,0,1)--cycle;
\fill[zcol!26,draw=gray!80,fill opacity=0.9] (1,0,0)--(1,1,0)--(1,1,1)--(1,0,1)--cycle;
\fill[zcol!26,draw=gray!80,fill opacity=0.9] (0,0,1)--(1,0,1)--(1,1,1)--(0,1,1)--cycle;
\node at (0.5,0,0.5) {\small$Z$};
\node at (1,0.5,0.5) {\small$Z$};
\node at (0.5,0.5,1) {\small$Z$};
\node at (0.5,0,-0.55) {(b) $Z$-check ${\cal S}^Z_{s_3}$};
\end{scope}
\begin{scope}[xshift=5.4cm,yshift=0.05cm,scale=0.85]
\foreach \i in {0,...,4} {\draw[gray!35] (\i,0,0)--(\i,4,0); \draw[gray!35] (0,\i,0)--(4,\i,0);}
\fill[zcol!26,draw=gray!80] (1,1,0)--(3,1,0)--(3,2,0)--(2,2,0)--(2,3,0)--(1,3,0)--cycle;
\draw[gray!80] (2,1,0)--(2,2,0) (1,2,0)--(2,2,0);
\draw[xcol,line width=1.6pt] (1,1,0)--(3,1,0)--(3,2,0)--(2,2,0)--(2,3,0)--(1,3,0)--cycle;
\node at (1.5,1.5,0) {\small$Z$};
\node at (2.5,1.5,0) {\small$Z$};
\node at (1.5,2.5,0) {\small$Z$};
\node[xcol,right] at (3.05,1.5,0) {$\partial\Sigma_2$};
\node[zcol!70!black] at (0.45,2.6,0) {$\Sigma_2$};
\node at (2.6,0,-0.65) {(c) $U^e(\Sigma_2)$ and its loop excitation};
\end{scope}
\end{tikzpicture}
\caption{\label{fig:geometry}Geometry of the 2-form code, drawn in a three-dimensional slice of a hypercubic lattice. Qubits sit on the 2-cells (plaquettes). (a) The $X$-check ${\cal S}^X_e$ acts on the 2-cells that contain the edge $e$: four in this slice, and $2(2N-1)$ on a $2N$-dimensional hypercubic lattice. (b) The $Z$-check ${\cal S}^Z_{s_3}$ acts on the six 2-cells in the boundary of the 3-cell $s_3$. An $X$-check and a $Z$-check share zero or two 2-cells, so they commute. (c) The electric operator $U^e(\Sigma_2)=\prod_{f\in\Sigma_2}Z_f$ on an open surface. An $X$-check on an interior edge meets two of its 2-cells and is not violated; an $X$-check on a boundary edge meets one and is violated. The excitation is therefore the loop $\partial\Sigma_2$ (red). If $\Sigma_2$ is closed there is no excitation, and $U^e(\Sigma_2)$ is a logical $\bar Z$ when $\Sigma_2$ is homologically nontrivial.}
\end{figure*}
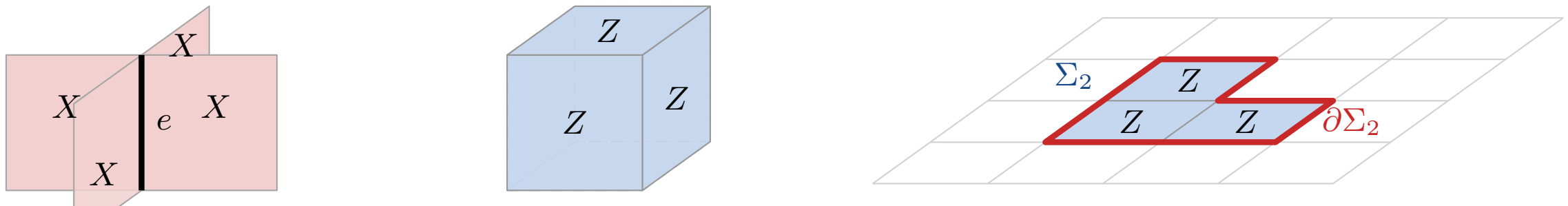

\paragraph{Definition.}
Take any cellulation (for example a triangulation) of a closed $2N$-dimensional manifold $M$, and place one physical qubit on every 2-cell $f$. The stabilizers are
\begin{equation}
    {\cal S}^X_{e}=\prod_{f\supset e} X_f,\qquad {\cal S}^Z_{s_3}=\prod_{f\subset \partial s_3} Z_f~,
\end{equation}
one $X$-check for every edge $e$, acting on the 2-cells that contain $e$, and one $Z$-check for every 3-cell $s_3$, acting on the 2-cells in its boundary (Fig.~\ref{fig:geometry}(a,b)). 
This is the homological (CSS) code of the $\mathbb{Z}_2$ chain complex $C_3\xrightarrow{\partial_3}C_2\xrightarrow{\partial_2}C_1$, with $H_Z=\partial_3^{T}$ and $H_X=\partial_2$. We call it the 2-form $\mathbb{CP}^N$ code when $M=\mathbb{CP}^N$. Here ``2-form'' records that the qubits live on 2-cells, so that the code is the lattice $\mathbb{Z}_2$ 2-form gauge theory; this distinguishes it from other homological codes on $\mathbb{CP}^N$ with qubits on cells of other dimensions. (On tori this is the familiar loop toric code; since $\mathbb{CP}^N$ is not a torus we use the term homological code.)

Table~\ref{tab:dictionary} collects the standard dictionary between the code and the topology of $M$, which we use throughout. A $Z$-type logical operator is a product of $Z$ over a closed surface, i.e.\ a 2-cycle, that is not the boundary of a 3-chain. An $X$-type logical operator is a product of $X$ over the 2-cells dual to a closed $(2N-2)$-dimensional hypersurface, i.e.\ a 2-cocycle that is not a coboundary. The number of logical qubits is therefore $k=\dim H_2(M;\mathbb{Z}_2)$. The $Z$-distance $d_Z$ is the smallest number of 2-cells in a homologically nontrivial closed surface, and the $X$-distance $d_X$ is the smallest number of 2-cells crossed by a nontrivial $(2N-2)$-dimensional hypersurface.

\begin{table}[t]
\caption{\label{tab:dictionary}Dictionary between the 2-form homological code and the cellulation of a $2N$-manifold $M$.}
\begin{ruledtabular}
\begin{tabular}{ll}
cells of $M$ & role in the code\\
\hline
0-cells (vertices) & $X$-metachecks (check redundancies)\\
1-cells (edges) & $X$-checks ${\cal S}^X_e$\\
2-cells & physical qubits\\
3-cells & $Z$-checks ${\cal S}^Z_{s_3}$\\
4-cells & $Z$-metachecks (check redundancies)\\
\hline
nontrivial 2-cycles & $Z$-type logical operators\\
nontrivial 2-cocycles & $X$-type logical operators\\
$\dim H_2(M;\mathbb{Z}_2)$ & number of logical qubits $k$\\
\end{tabular}
\end{ruledtabular}
\end{table}

\paragraph{Electric and magnetic operators.} The $Z$-type and $X$-type operators of the code have a direct physical interpretation in the $\mathbb{Z}_2$ 2-form gauge theory. The electric operator is a product of $Z$ over a surface $\Sigma_2$ made of 2-cells, and the magnetic operator is a product of $X$ over the 2-cells crossed by a $(2N-2)$-dimensional hypersurface $\Sigma_{2N-2}$ of the dual lattice:
\begin{equation}
    U^e(\Sigma_2)=\prod_{f\in\Sigma_2} Z_f,\quad V^m(\Sigma_{2N-2})=\prod_{f\cap \Sigma_{2N-2}\neq 0} X_f~.
\end{equation}
If $\Sigma_2$ has a boundary, $U^e$ violates the ${\cal S}^X$ checks along $\partial\Sigma_2$, creating a loop-like electric excitation (Fig.~\ref{fig:geometry}(c)); if $\Sigma_{2N-2}$ has a boundary, $V^m$ violates the ${\cal S}^Z$ checks along it, creating a $(2N-3)$-dimensional magnetic excitation. The two operators commute up to the parity of the number of intersection points of $\Sigma_2$ and $\Sigma_{2N-2}$ in the $2N$-dimensional space:
\begin{equation}
\begin{split}
    U^e(\Sigma_2)V^m(\Sigma_{2N-2}')&=V^m(\Sigma_{2N-2}')U^e(\Sigma_2)\\
    &\quad\times(-1)^{\#(\Sigma_2,\Sigma'_{2N-2})}~.
\end{split}
\end{equation}
For closed $\Sigma_2$ and $\Sigma_{2N-2}$ the operators commute with all checks; they are logical operators when the surfaces are homologically nontrivial, and stabilizers otherwise.

\paragraph{Single logical qubit.} When $\Sigma_2$ and $\Sigma_{2N-2}$ are homologically nontrivial, the electric and magnetic operators are the logical $\bar Z$ and $\bar X$. Since $H_2(\mathbb{CP}^N;\mathbb{Z}_2)=\mathbb{Z}_2$, there is a single nontrivial class: the nontrivial 2-cycle is a projective line $\mathbb{CP}^1\subset\mathbb{CP}^N$ and the nontrivial $(2N-2)$-cycle is a hyperplane $\mathbb{CP}^{N-1}\subset\mathbb{CP}^N$. They intersect in a single point, so $\bar X$ and $\bar Z$ anticommute and the code has exactly one logical qubit for every $N$. By contrast, the torus $T^{2N}$ gives $k=\binom{2N}{2}$, e.g.\ $k=15$ for $T^6$. This single logical qubit is the main reason for working on $\mathbb{CP}^N$.

Equivalently, code states are labelled by the flat $\mathbb{Z}_2$ 2-form gauge fields $a$ (2-cocycles) up to gauge transformations, i.e.\ by $H^2(\mathbb{CP}^N;\mathbb{Z}_2)=\mathbb{Z}_2$. The nontrivial class is the mod-2 reduction of the K\"ahler form $\omega$, so the logical basis states are $a=n\,\omega$ with $n=0,1$.

\paragraph{Stabilizer weights.} The weights of the stabilizers on a hypercubic lattice are as follows. Denote the space directions by $x^1,x^2,\cdots, x^{2N}$. An edge in direction $x^1$ is adjacent to faces in the $x^1-x^i$ directions for $i=2,3,\cdots,2N$, and the $x^i$ direction can be either positive or negative. This gives $2(2N-1)$ faces adjacent to each edge.
Thus the $X$-stabilizer ${\cal S}^X$ has weight $w_X=4N-2$.
A cube has 6 faces on the boundary, and thus the $Z$-stabilizer has weight $w_Z=6$ independent of $N$. 

Since $\mathbb{CP}^N$ is not a torus, it admits no globally hypercubic lattice, and the counts above hold only in regions where the cellulation is locally hypercubic. On a general cellulation the $Z$-stabilizer weight is still fixed by the cell type (e.g., $w_Z=6$ on any cubical complex and $w_Z=4$ on any triangulation), whereas the $X$-stabilizer weight of an edge $e$ equals the number of 2-cells containing $e$, i.e.\ the number of vertices of the link of $e$, which varies from edge to edge. For instance, on the 18-vertex triangulation of $\mathbb{CP}^3$ studied in Sec.~\ref{sec:practical}, the $X$-stabilizers have weights between 8 and 16. The statement that only $w_X$ grows with $N$ is therefore about the typical (locally hypercubic) weight.

We note that the weights can also be reduced to $O(1)$ using the methods in \cite{Hastings:2016dtg,Hastings:2021kfa}, but the compatibility with low-depth logical gate constructions is less clear in that formalism.

\paragraph{Distance scaling.} For a lattice of linear size $L$ there are $n\sim L^{2N}$ qubits, a $Z$-logical is a closed surface of area $\sim L^2$, and an $X$-logical is a closed hypersurface of size $\sim L^{2N-2}$. The distance is therefore $d=d_Z=O(n^{1/N})$. Exact code parameters in the 2-form $\mathbb{CP}^3$ code on explicit examples of triangulations are described in Sec.~\ref{sec:practical}.

\section{Self-Correction and Single-Shot Error Correction}
\label{sec:memory}

For $N\ge2$ the 2-form code has only extended excitations: the electric excitation is a loop, and the magnetic excitation has dimension $2N-3\ge1$. This gives the code two properties that the 2D surface code lacks. It is a self-correcting quantum memory, which we prove in Sec.~\ref{sec:selfcorrection}. Its syndromes also obey local consistency conditions that allow single-shot error correction, which we describe in Sec.~\ref{sec:singleshot}.

\subsection{Self-correcting quantum memory}
\label{sec:selfcorrection}

Because the excitations are extended, creating a logical error requires growing a large membrane, whose energy cost grows with its boundary. This is the mechanism of self-correction. For the 4D toric code, the resulting energy barrier gives a memory lifetime that grows exponentially with system size below a critical temperature under weak thermal coupling to a bath~\cite{Dennis:2001nw,doi:10.1142/S1230161210000023}; see also Ref.~\cite{Bombin_2013} for the 6D color code. We prove the same for the 2-form $\mathbb{CP}^N$ code.

Self-correction is a statement about a family of codes with growing system size. The family must keep the number of simplices that meet at a vertex uniformly bounded. Repeated barycentric subdivision does not do this, but edgewise subdivision (i.e., subdividing each simplex into a fine mesh by hyperplanes in a uniform fashion) does. For a triangulation $K$ with ordered vertices, the edgewise subdivision $K_m$ cuts every simplex of $K$ into simplices of finitely many shapes, $m^{2N}$ of them in each top simplex (Appendix~\ref{app:sc-lattice}). 

\begin{theorem}[Self-correcting quantum memory]\label{thm:selfcorrection}
Let $N\ge2$, let $K$ be a triangulation of $\mathbb{CP}^N$ with ordered vertices, and fix an integer $q>6(2N+1)$. For $m\in q\mathbb{N}$, consider the 2-form code on the edgewise subdivision $K_m$, which has $n=O(m^{2N})$ qubits. Couple its Hamiltonian $H=-J\sum_e{\cal S}^X_e-J\sum_{s_3}{\cal S}^Z_{s_3}$ weakly to a thermal bath at inverse temperature $\beta$ through single-qubit Pauli operators, with Davies dynamics $e^{t\mathcal{L}}$ (Definition~\ref{def:sc-bath}). There are constants $\beta_0$ and $C$, which depend on $K$, $q$ and $J$ but not on $m$, such that for all $\beta\ge\beta_0$, $m\in q\mathbb{N}$, $t\ge0$ and logical states $\bar\rho$,
\begin{equation}\label{eq:selfcorrection}
\begin{split}
&\big\|\mathcal{D}_m\big(e^{t\mathcal{L}}\rho_\beta(\bar\rho)\big)-\bar\rho\big\|_1\\
&\qquad\le C\,m^{4N}\,(1+h_{\max}t)\,2^{-m/(2q(2N+1))}~.
\end{split}
\end{equation}
Here $\rho_\beta(\bar\rho)$ is the thermal encoding of $\bar\rho$ (Definition~\ref{def:sc-encoding}), $\mathcal{D}_m$ is the local-filling decoder (Definition~\ref{def:sc-decoder}), and $h_{\max}$ is the largest transition rate of the bath.
\end{theorem}

The theorem starts from the thermal encoding $\rho_\beta(\bar\rho)$, not from an exact code state. It is $\bar\rho$ with a Pauli error applied, drawn from the Gibbs distribution of errors conditioned on the trivial logical class of the decoder; for the maximally mixed logical state the four classes together give the Gibbs state of $H$. As in Ref.~\cite{doi:10.1142/S1230161210000023}, it is an idealized initial state for the theorem. 
Starting with this initial thermal encoding, the theorem gives an exponentially growing lower bound on the memory time, the time after which the error~\eqref{eq:selfcorrection} reaches a fixed value $\epsilon$: the memory time $\tau_\epsilon(m)$ is at least $e^{\Omega(m)}=e^{\Omega(n^{1/2N})}$. 
The proof follows Ref.~\cite{doi:10.1142/S1230161210000023} and detailed description is relegated to Appendix~\ref{app:selfcorrection}.

\subsection{Single-shot error correction via metachecks}
\label{sec:singleshot}

The extended excitations also motivate single-shot error correction~\cite{Bombin:2014ksv}. The syndrome of a data error must itself form closed loops (or closed $(2N-3)$-dimensional objects); a measurement error typically breaks this closure and is detected by local consistency checks, which are the metachecks described below. One round of noisy syndrome measurement is then expected to suffice to correct data and measurement errors up to a small, local residual error, in contrast with the $O(d)$ repeated rounds needed by codes such as the 2D surface code.

Single-shot error correction can be implemented using metachecks~\cite{Bombin:2014ksv,Campbell:2019cyo}, which are ``checks on the checks.'' They exist because certain products of checks are the identity operator. A noiseless syndrome must respect these relations, and a violated relation signals a measurement error. The metachecks are associated with the 4-cells (for $Z$-checks) and 0-cells (for $X$-checks) of the lattice.
For every 4-cell, the product of the $Z$-checks on the 3-cells on its boundary is the identity:
\begin{equation}
    \prod_{s_3\in \partial s_4} S^Z_{s_3}=1~.
\end{equation}
Likewise, for every 0-cell, the product of the $X$-checks on the edges that contain it is the identity, since every 2-cell containing the 0-cell contains exactly two of those edges:
\begin{equation}
    \prod_{s_1:s_0\in \partial s_1} S^X_{s_1}=1~.
\end{equation}
If the measured syndrome violates one of these relations, a measurement error has occurred, and the decoder can account for it within the same round. On the explicit lattices of Sec.~\ref{sec:practical} the metachecks are abundant: the $[[783,1,4]]$ code has $3\,021$ weight-5 metachecks on its 4-simplices and $18$ on its vertices, and the decoder of the one-round simulations of Sec.~\ref{sec:simulation} uses them. Metachecks do not detect every measurement error: faulty outcomes that themselves form a closed syndrome satisfy all metachecks and look like a data error. Single-shot decoding does not need to tell them apart; it needs the residual error after each round to stay small and local, so that it does not accumulate. For a rigorous proof for single-shot decoding~\cite{Bombin:2014ksv,Campbell:2019cyo} it is necessary to establish such locality of residual errors; we have not established such rigorous results for the lattices of this paper, and the simulations of Sec.~\ref{sec:simulation} test a single round.

\begin{figure}[t]
\centering
\begin{tikzpicture}[line join=round]
\begin{scope}
\coordinate (a) at (0,0); \coordinate (b) at (1.8,0); \coordinate (c) at (0.9,1.45);
\fill[zcol!26] (a)--(b)--(c)--cycle;
\node at (0.9,0.5) {\small$Z$};
\draw[xcol,line width=1.6pt] (a)--(b)--(c)--cycle;
\foreach \p in {a,b,c} \draw[fill=white] (\p) circle (2.3pt);
\node[below left] at (a) {\small$v_0$}; \node[below right] at (b) {\small$v_1$}; \node[above] at (c) {\small$v_2$};
\node[align=center,font=\footnotesize] at (0.9,-0.65) {(a) data error:\\ closed loop};
\end{scope}
\begin{scope}[xshift=3.6cm]
\coordinate (a) at (0,0); \coordinate (b) at (1.8,0); \coordinate (c) at (0.9,1.45);
\draw[gray!60] (a)--(c)--(b);
\draw[xcol,line width=1.6pt] (a)--(b);
\draw[fill=xcol] (a) circle (2.3pt); \draw[fill=xcol] (b) circle (2.3pt); \draw[fill=white] (c) circle (2.3pt);
\node[below left] at (a) {\small$v_0$}; \node[below right] at (b) {\small$v_1$}; \node[above] at (c) {\small$v_2$};
\node[align=center,font=\footnotesize] at (0.9,-0.65) {(b) measurement error:\\ open path};
\end{scope}
\end{tikzpicture}
\caption{\label{fig:metacheck}Metachecks on a single triangle. Red edges
are violated $X$-checks, and a filled vertex is a violated metacheck (an
odd number of violated edges at that vertex). (a)~A $Z$ error on the
triangle violates the checks on its three edges; the syndrome is a closed
loop and no metacheck fires. (b)~A faulty measurement of the check on
$v_0v_1$ alone gives an open path, and the metachecks at its endpoints
fire.}
\end{figure}
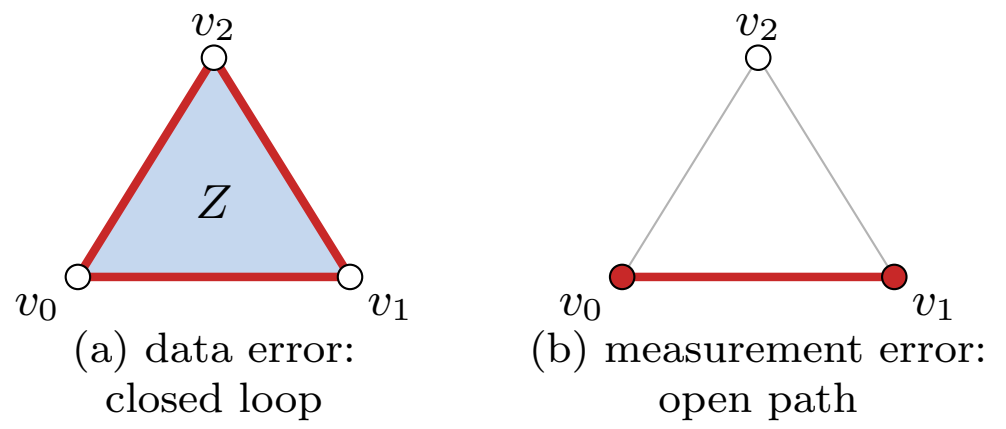

Figure~\ref{fig:metacheck} shows the simplest case. A $Z$ error on a triangle $v_0v_1v_2$ violates the $X$-checks on its three edges. The syndrome is the closed loop $\partial(v_0v_1v_2)$, each of its vertices meets two violated edges, and every metacheck is satisfied. If instead only the measurement of the check on $v_0v_1$ fails, the syndrome is that single edge, then the metachecks at $v_0$ and $v_1$ fail. Data errors always form closed syndromes and measurement errors break them open, whose endpoints violates the metacheck. The $Z$-checks and the metachecks on 4-cells work in the same way with one dimension higher.

\section{Logical Gates and Universal Quantum Computation}
\label{sec:gates}

In this section we construct Clifford gates in the 2-form $\mathbb{CP}^2$ code and a $CCZ$ gate in the 2-form $\mathbb{CP}^3$ code, implemented by constant-depth local unitaries, using the method of Ref.~\cite{Hsin:2024nwc}, which generalizes the constructions in e.g.~\cite{Kubica:2015mta,Barkeshli:2022edm,Barkeshli:2022wuz,Barkeshli:2023bta,Hsin:2024pdi}. We then propose a universal gate set built from these Clifford gates and the magic states obtained by switching from the 2-form $\mathbb{CP}^3$ code to the 2-form $\mathbb{CP}^2$ code. Table~\ref{tab:gates} summarizes the gates.

\paragraph{Locality and depth.} We separate three statements for each
gate: its logical action, which holds on any cellulation on which the
cohomology operation is defined (Theorems~\ref{thm:CZ}--\ref{thm:CCZ});
its locality; and its depth on a given cellulation. Depth counts layers
of the physical gates. Each circuit is a product of commuting diagonal
gates, each supported in one top cell (for $HSH$ on a triangulation, on
the triangles at one vertex; Sec.~\ref{sec:hshgate}). We say that a
family of cellulations has \textit{bounded incidence} if every 2-cell
lies in at most $\Delta$ top cells and every vertex in at most $\Delta$
cells, with $\Delta$ independent of the system size, as for the
edgewise subdivisions of Sec.~\ref{sec:selfcorrection}. 

This property about the cellulation is required to bound the depth of the circuit for the logical operator.
If the cellulation has bounded incidence, every
qubit takes part in at most $g=O(\Delta)$ gates of at most $\kappa=O(\Delta)$
qubits each, and a greedy colouring schedules them in depth at most
$\kappa(g-1)+1$; since the gates commute, every such schedule implements the
same unitary. 
On a triangulation, the depth of the circuit is at least the maximal
number of top simplices in which one 2-cell lies on; on an explicit triangulation 
$\mathbb{CP}^3_{18}$ this bound is $27$ for the logical $CCZ$ gate
$U_{\rm CCZ}$, and a greedy schedule of its $622$ CCZ gates saturates this lower bound
(Sec.~\ref{sec:cczprep}).

\begin{table}[t]
\caption{\label{tab:gates}Logical gates of the 2-form codes implemented
by constant-depth local unitaries (Sec.~\ref{sec:gates}).
4D: the 2-form $\mathbb{CP}^2$ code; 6D: the 2-form $\mathbb{CP}^3$
code; $\times2$ and $\times3$: number of copies on the same lattice. Each
circuit multiplies a configuration $a$ of the $Z$ basis (or $a^\vee$ of
the $X$ basis) by the phase shown, which depends only on the logical
label. Every circuit has bounded depth on a family with bounded
incidence. The depth column gives reference depths for hypercubic
lattices, which $\mathbb{CP}^N$ does not admit globally; on the
triangulation $\mathbb{CP}^3_{18}$ of Sec.~\ref{sec:practical} a complete
schedule of the $CCZ$ circuit has depth $27$. As for any CSS code, two copies also have a logical CNOT
applied qubit by qubit.}
\begin{ruledtabular}
\begin{tabular}{lllll}
gate & code & circuit & \begin{tabular}[b]{@{}l@{}}hypercubic\\ reference depth\end{tabular} & Sec.\\
\hline
$CZ$ & 4D $\times2$ & $(-1)^{\int a_1\cup a_2}$ & $2$ & \ref{sec:czgate}\\
$S$ & 4D & $i^{\int{\cal P}(a)}$ & $2$ & \ref{sec:sgate}\\
$HSH$ & 4D & $i^{\int{\cal P}(a^\vee)}$ & $2$ & \ref{sec:hshgate}\\
$H$ & 4D & $U_\text{S}\,U_\text{HSH}\,U_\text{S}$ & $\le6$ & \ref{sec:hgate}\\
$CCZ$ & 6D $\times3$ & $(-1)^{\int a_1\cup a_2\cup a_3}$ & $3$ & \ref{sec:ccz}\\
\end{tabular}
\end{ruledtabular}
\end{table}

Let us outline the construction of the logical gates. In the $Z$ basis, a configuration of the qubits is a $\mathbb{Z}_2$-valued 2-cochain $a$, with $a(f)=(1-Z_f)/2$, and code states are superpositions of 2-cocycles. A diagonal circuit that multiplies each configuration by a phase $e^{i\int_M F(a)}$, where $F$ is a cohomology operation such as a cup product, is a logical gate. 

Here the integral of a top cochain $c$ is its evaluation on the fundamental chain; on an oriented triangulation,
\begin{equation}\label{eq:fundchain}
\int_M c=\sum_{s_d}\epsilon(s_d)\,c(s_d)~,
\end{equation}
where  $\epsilon(s_d)=+1$ if the vertex order of the top simplex $s_d$ agrees with the orientation of $M$ and $-1$ otherwise. 
For a cocycle $a$, $\int_M F(a)$ is unchanged by gauge transformations $a\to a+\delta\lambda$, so it depends only on the logical label $n$ in $a=n\,\omega$ (Sec.~\ref{sec:CPN}). 

\begin{figure}[t]
\centering
\begin{tikzpicture}[scale=1.0,line join=round]
\begin{scope}
\foreach \k in {0,...,4} \coordinate (p\k) at ({198-72*\k}:1.15);
\foreach \i in {0,...,4} \foreach \j in {0,...,4} {\ifnum\i<\j \draw[gray!40] (p\i)--(p\j); \fi}
\fill[cupA,fill opacity=0.45,draw=cupA!80!black,thick] (p0)--(p1)--(p2)--cycle;
\fill[cupB,fill opacity=0.45,draw=cupB!80!black,thick] (p2)--(p3)--(p4)--cycle;
\foreach \k in {0,...,4} {\fill (p\k) circle (1.3pt); \node at ({198-72*\k}:1.45) {\small$v_\k$};}
\node[cupA!70!black] at ($(p0)!0.5!(p1)!0.33!(p2)$) {\small$a_1$};
\node[cupB!70!black] at ($(p3)!0.5!(p4)!0.33!(p2)$) {\small$a_2$};
\node at (0,-1.85) {(a) $a_1\cup a_2$: one $CZ$};
\end{scope}
\begin{scope}[xshift=4.3cm]
\foreach \k in {0,...,6} \coordinate (q\k) at ({90+180/7+360/7*\k}:1.15);
\foreach \i in {0,...,6} \foreach \j in {0,...,6} {\ifnum\i<\j \draw[gray!30] (q\i)--(q\j); \fi}
\fill[cupA,fill opacity=0.45,draw=cupA!80!black,thick] (q0)--(q1)--(q2)--cycle;
\fill[cupB,fill opacity=0.45,draw=cupB!80!black,thick] (q2)--(q3)--(q4)--cycle;
\fill[cupC,fill opacity=0.45,draw=cupC!80!black,thick] (q4)--(q5)--(q6)--cycle;
\foreach \k in {0,...,6} {\fill (q\k) circle (1.3pt); \node at ({90+180/7+360/7*\k}:1.45) {\small$v_\k$};}
\node at (0,-1.85) {(b) $a_1\cup a_2\cup a_3$: one $CCZ$};
\end{scope}
\end{tikzpicture}
\caption{\label{fig:cup}Cup-product gates on a single top simplex with ordered vertices $v_0<v_1<\cdots$. (a) On a 4-simplex, $a_1\cup a_2$ pairs the front triangle $v_0v_1v_2$ (copy 1) with the back triangle $v_2v_3v_4$ (copy 2), which share only the vertex $v_2$. The circuit $(-1)^{\int a_1\cup a_2}$ applies one physical $CZ$ between these two qubits for every 4-simplex. (b) On a 6-simplex, $a_1\cup a_2\cup a_3$ uses the front $v_0v_1v_2$, middle $v_2v_3v_4$ and back $v_4v_5v_6$ triangles of copies 1, 2 and 3, and $(-1)^{\int a_1\cup a_2\cup a_3}$ applies one physical $CCZ$ for every 6-simplex.}
\end{figure}
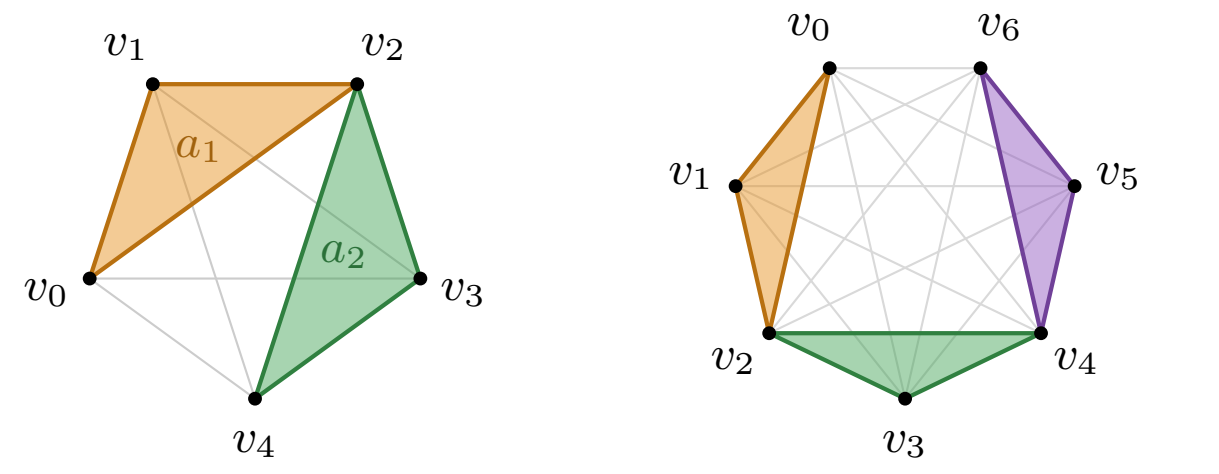

The circuit representation of the logical operator is read from the cochain level formula for the cohomology operation. For instance, on a triangulation with ordered vertices, the cup product of two 2-cochains evaluated on a 4-simplex $(v_0v_1v_2v_3v_4)$ is
\begin{equation}
    (a\cup b)(v_0v_1v_2v_3v_4)=a(v_0v_1v_2)\,b(v_2v_3v_4)~,
\end{equation}
the product of $a$ on the ``front'' triangle and $b$ on the ``back'' triangle (Fig.~\ref{fig:cup}). Consequently $(-1)^{\int a_1\cup a_2}$ is a product of physical $CZ$ gates, one for each 4-simplex, between the front triangle of the first copy and the back triangle of the second. 
Likewise $(-1)^{\int a_1\cup a_2\cup a_3}$ is a product of physical $CCZ$ gates, one for each 6-simplex, acting on its front, middle and back triangles. 

To see why the cup product gives a logical operator, take e.g., $F=a_1\cup a_2$ and apply a gauge transformation $a_1\to a_1+\delta\lambda$ to the first copy. This gauge transformation corresponds to the conjugation action of a $X$ stabilizer on the cochain $a$. Since $a_2$ is a cocycle within the $Z$ stabilizer subspace, the Leibniz rule gives $\delta\lambda\cup a_2=\delta(\lambda\cup a_2)$, and the integral of a coboundary over a closed manifold vanishes, so
\begin{equation}
\int_M(a_1+\delta\lambda)\cup a_2=\int_M a_1\cup a_2+\int_M\delta(\lambda\cup a_2)=\int_M a_1\cup a_2~,
\end{equation}
which implies that this cup-product operator $(-1)^{\int a_1\cup a_2}$ commutes with the $X$ stabilizers within the $Z$ stabilizer subspace satisfying $\delta a_1=\delta a_2=0$, therefore becomes a logical gate. More generally, an operator formed by a cohomology operation $F(a)$ is gauge invariant and defines a logical operator.

The logical action is read off from the cohomology ring of $M$.
Since $\int_{\mathbb{CP}^2}\omega^2=1$ and $\int_{\mathbb{CP}^3}\omega^3=1$, the circuits $(-1)^{\int a_1\cup a_2}$, $(-1)^{\int a_1\cup a_2\cup a_3}$ implement logical $CZ$ and $CCZ$ gates, respectively.

\subsection{Diagonal and non-diagonal logical gates from cohomology invariants}

Let us begin with a brief review of the construction in \cite{Hsin:2024nwc} for logical gates implemented by constant-depth local unitaries. The construction gives diagonal logical gates as well as non-diagonal logical gates that are diagonal in the $X$-basis. We emphasize that both types of gates (diagonal in the $Z$-basis and diagonal in the $X$-basis) can exist in the same code.

\subsubsection{Diagonal logical gates in $Z$-basis}
\label{sec:cohomologyZbasis}

Consider physical qubits on $q$-cells (we focus on $q=2$), so that the gauge field is a $q$-form gauge field.
For an $r$-cocycle $\omega\in H^r(B^q\mathbb{Z}_2,U(1))$ and an $r$-dimensional closed subspace $M_r$, we can construct logical gate by integrating $\omega$ over the $r$-dimensional subspace:
\begin{equation}
    U(\omega,M_r)=e^{i\int_{M_r} \omega(a)}~,
\end{equation}
which is a constant-depth local unitary that acts on the Hilbert space as
\begin{equation}
    U(\omega,M_r)|\{a(s_q)\}\rangle=\prod_{s_r\in M_r} e^{i \omega(s_r)}|\{a(s_q)\}\rangle~,
\end{equation}
where $|\{a(s_q)\}\rangle$ is the configuration of all physical qubits in the computational basis $(-1)^{a(s_q)}=Z_{s_q}$ on different $q$-cells $s_q$, and
$\omega(s_r)$ is the evaluation of $\omega$ on the $r$-cell $s_r$ for the physical qubit configuration $a(s_{q}')$ for the $q$-cells $s_{q}'\in s_r$. The depth of $U(\omega,M_r)$ depends on the cocycle $\omega$ and on the cellulation.

While here we focus on single physical qubit on every $q$-cell, it is easy to generalize to multiple types of physical qubits on cells of different dimensions. For physical qubit of type $i$ on $q_i$-dimensional cell, the cocycle is now $\omega\in H^r({\cal B},U(1))$ for ${\cal B}=\prod_i B^{q_i} \mathbb{Z}_2^{(i)}$ where the superscript in $\mathbb{Z}_2^{(i)}=\mathbb{Z}_2$ denotes the $i^\text{th}$ type of physical qubits.

\subsubsection{Diagonal logical gates in $X$-basis: non-diagonal gates}
\label{sec:cohomologyXbasis}

In addition to the diagonal gates in $Z$-basis, we can also consider diagonal gates in the $X$-basis using the dual gauge field $a^\vee$.
Denote the degree of the dual gauge field by $q^\vee=D-1-q$, where $D=2N+1$ is the spacetime dimension, so that $q^\vee=2N-q$ is the degree of the dual field on the dual cellulation of space; for the 2-form $\mathbb{CP}^2$ code, $q^\vee=2$. 

Given an $r$-cocycle $\omega^\vee\in H^{r}(B^{q^\vee}\mathbb{Z}_2,U(1))$ and an $r$-dimensional closed subspace $M_{r}^\vee$ of the dual lattice, we can construct a logical gate by integrating $\omega^\vee$ over that subspace:
\begin{equation}
    U(\omega^\vee,M_{r}^\vee)=e^{i\int_{M_{r}^\vee} \omega^\vee(a^\vee)}~,
\end{equation}
which is a constant-depth local unitary that acts on the Hilbert space as 
\begin{equation}
    U(\omega^\vee,M_{r}^\vee)=\prod_{s_r^\vee\in M_r^\vee} e^{i \omega^\vee(s_r^\vee)}~,
\end{equation}
where $\omega^\vee(s_r^\vee)$ is
the evaluation of $\omega^\vee$ on the $r$-cell $s_r^\vee$. The evaluation depends on the values $a^\vee(s_{q^\vee}^\vee)$ for $s_{q^\vee}^\vee\subset s_r^\vee$, defined as
\begin{equation}
    a^\vee(s_{q^\vee}^\vee)=\frac{1-X_{s_q}}{2},\quad \#(s_q,s_{q^\vee}^\vee)=1~.
\end{equation}
In other words, $a^\vee(s_{q^\vee}^\vee)$ is given by $X$ on the cell $s_q$ of the original lattice dual to $s_{q^\vee}^\vee$ on the dual lattice.
As in the $Z$ basis, the depth depends on the cocycle $\omega^\vee$ and on the cellulation.

For example, for $q=2$, $r=q^\vee$, and $\omega^\vee$ is the generator of $H^{q^\vee}(B^{q^\vee}\mathbb{Z}_2,U(1))=\mathbb{Z}_2$, this operator is the magnetic logical operator
\begin{align}
    &U(\omega^\vee,M_{q^\vee}^\vee)=\prod_{s_{q^\vee}^\vee\in M_{q^\vee}^\vee} (-1)^{a^\vee(s_{q^\vee}^\vee)}\cr
    &=\prod_{s^\vee_{q^\vee}\in M_{q^\vee}^\vee}\prod_{s_q:\#(s_q,s_{q^\vee}^\vee)=1} (-1)^{(1-X_{s_q})/2} \cr 
    &=\prod_{s_q:\#(s_q,M_{q^\vee}^\vee)=1} X_{s_q}~.
\end{align}
Similarly, we can also generalize the expression to multiple types of physical qubits on cells of different dimensions.

\subsection{Constant-depth local unitary CZ gate in 2-form $\mathbb{CP}^2$ code}\label{sec:czgate}

Consider two copies of 2-form $\mathbb{CP}^2$ code. Denote the gauge fields for the two copies by $a_1,a_2$, where $a_i(s_2)=(1-Z^i_{s_2})/2$ for physical Pauli operators $Z^i_{s_2}$ with $i=1,2$ on 2-cell $s_2$. Consider the cocycle $\omega=a_1\cup a_2$. Integrating the cocycle over $M_4=\mathbb{CP}^2$ gives the logical gate
\begin{align}
  &U_\text{CZ}=(-1)^{\int_{\mathbb{CP}^2}a_1\cup a_2}\cr 
  &=\prod_{s_4\in M_4} \prod_{s_2,s_2'\in s_4:\; c_{s_4}(s_2,s_2')\neq0}(-1)^{a_1(s_2)a_2(s_2')}
~.
\end{align}
On a hypercubic lattice this circuit has depth 2.

\begin{theorem}\label{thm:CZ}
    On a triangulation of $\mathbb{CP}^2$ with ordered vertices, or any cellulation on which the cup product is defined, the circuit $U_\text{CZ}$ implements the logical CZ gate between two copies of the 2-form $\mathbb{CP}^2$ code. It is a constant-depth local unitary on any family of cellulations with bounded incidence.
\end{theorem}
\begin{proof}
    The phase $(-1)^{\int a_1\cup a_2}$ depends only on the cohomology classes of $a_1$ and $a_2$, by the gauge-invariance argument of Sec.~\ref{sec:gates}, so it suffices to evaluate it on one representative of each logical state. We label the two logical qubits for the two copies by the holonomies $a_1=n_1 \omega$, $a_2=n_2\omega$ for $n_1,n_2=0,1$. The logical state can be labeled by $|n_1,n_2\rangle$.
The circuit acts on the logical state as
\begin{equation}
    U_\text{CZ}|n_1,n_2\rangle=(-1)^{n_1n_2}|n_1,n_2\rangle=\overline{\text{CZ}}_{n_1,n_2}|n_1,n_2\rangle~,
\end{equation}
which is the logical CZ gate for the two logical qubits.
\end{proof}

\subsection{Constant-depth local unitary S gate in 2-form $\mathbb{CP}^2$ code}\label{sec:sgate}

The S logical gate can be constructed using the Pontryagin square operation as in \cite{Hsin:2024pdi,Hsin:2024nwc}. 
Consider the 4-cocycle ${\cal P}(a)=a\cup a-a\cup_1 da$ for 2-cocycle $a$. Here $a$ is lifted to an integer cochain with values $0,1$, and the correction term $a\cup_1 da$, built from Steenrod's $\cup_1$ product, makes $\int{\cal P}(a)$ well defined modulo 4 rather than only modulo 2; this is what allows the phase $i^{\int{\cal P}(a)}$, and hence a logical $S$ rather than $Z$. On $\mathbb{CP}^2$, $\int{\cal P}(n\omega)=n^2\int\omega^2=n^2$ modulo 4.
Integrating the cocycle over $M_4=\mathbb{CP}^2$ gives the logical gate
\begin{align}
    &U_\text{S}(M_4)=i^{\int_{M_4} {\cal P}(a)}=i^{\int_{M_4} \left(a\cup a-a\cup_1 da\right)}\cr 
    &=\prod_{s_4\in M_4}\left(\prod_{s_2,s_2'\in s_4}i^{c_{s_4}(s_2,s_2')\,a(s_2)a(s_2')}\right)\cr
    &\quad \cdot\left(\prod_{\substack{s_2'',s_3\in s_4:\\ c^{(1)}_{s_4}(s_2'',s_3)\neq0}}\prod_{s_2'''\in \partial s_3}i^{o(s_2''',s_3)a(s_2'')a(s_2''')}\right)~.\label{eq:US}
\end{align}

Here, the coefficients $c$ are defined as follows. Let $\tilde s$ denote the indicator cochain of a cell $s$, then
\begin{equation}\label{eq:cupcoef}
c_{s_4}(s_2,s_2')=\epsilon(s_4)\,(\tilde s_2\cup\tilde s_2')(s_4)\in\{0,\pm1\}
\end{equation}
is the contribution of a top cell $s_4$ to $\int\tilde s_2\cup\tilde s_2'$; on a triangulation it equals $\epsilon(s_4)$ when $s_2$ and $s_2'$ are the front and back triangles of $s_4$, and vanishes otherwise. The coefficients $c_{s_6}(s_2,s_2',s_2'')$ of triple products and $c^{(1)}_{s_4}(s_2,s_3)=\epsilon(s_4)\,(\tilde s_2\cup_1\tilde s_3)(s_4)$ of $\cup_1$ products are defined in the same way.
On a triangulation, for $s_4=(v_0\cdots v_4)$ with $v_0<\dots<v_4$, $c^{(1)}_{s_4}$ is nonzero for the two pairs $(v_0v_3v_4,\,v_0v_1v_2v_3)$ and $(v_0v_1v_4,\,v_1v_2v_3v_4)$.

The sign $o(s_2''',s_3)=+ 1$ when $s_2'''$ appears in $\partial s_3$ with a $+$ sign, and $o(s_2''',s_3)=- 1$ when $s_2'''$ appears in $\partial s_3$ with a $-$ sign. In the above expression, for any 2-cell $s_2$, $a(s_2)=(1-Z_{s_2})/2$ for the physical Pauli $Z$ on $s_2$. In the first product, a factor is a controlled-$S$ gate when $c_{s_4}(s_2,s_2')=+1$ and a controlled-$S^\dagger$ gate when $c_{s_4}(s_2,s_2')=-1$. In the second, the sign of $c^{(1)}_{s_4}$ does not matter on code states, and a factor is a controlled-$S$ or controlled-$S^\dagger$ gate according to $o(s_2''',s_3)$. 

The last equality holds on code states with $da=0$ mod 2, and we take the product as the definition of the physical circuit $U_\text{S}$ on all configurations. On a configuration that satisfies the $Z$-checks, $da(s_3)=\sum_{s_2'''\in\partial s_3}o(s_2''',s_3)\,a(s_2''')$ is even, so $a(s_2'')\,da(s_3)$ is even and its sign does not matter modulo 4; the correction factors may therefore be written with exponent $+o(s_2''',s_3)\,a(s_2'')a(s_2''')$ as above, even though ${\cal P}(a)$ contains $-a\cup_1 da$. 

\begin{theorem}\label{thm:S}
    On a triangulation of $M_4=\mathbb{CP}^2$ with ordered vertices, oriented as in Eq.~\eqref{eq:fundchain}, the circuit $U_\text{S}$ implements the logical S gate of the 2-form $\mathbb{CP}^2$ code. It is a constant-depth local unitary on any family of triangulations with bounded incidence.
\end{theorem}
\begin{proof}
As for $U_\text{CZ}$, the phase $i^{\int{\cal P}(a)}$ depends only on the cohomology class of $a$, so it suffices to evaluate it on the logical labels. For $M_4=\mathbb{CP}^2$ and a single 2-form $\mathbb{CP}^2$ code, with logical qubit labeled by $a=n_a \omega$ for $n_a=0,1$ and $\omega$ being the K\"ahler form of $\mathbb{CP}^2$, this gives
\begin{equation}
    U_\text{S}|n_a\rangle=i^{n_a}|n_a\rangle=\overline{S}|n_a\rangle~.
\end{equation}
This is the logical S gate.    
\end{proof}

\subsection{Constant-depth local unitary HSH gate in 2-form $\mathbb{CP}^2$ code}\label{sec:hshgate}

For the construction of the Hadamard gate, it will be useful to construct the $HSH$ gate. The gate can also be obtained from the Pontryagin square operator as in the S gate construction. However, since conjugating by Hadamard changes the $Z$-basis diagonal gate into $X$-basis diagonal gate (i.e. non-diagonal in $Z$-basis), we will write the logical gate using the dual gauge field $a^\vee$ as in section \ref{sec:cohomologyXbasis}. 

The cocycle is $\omega^\vee(a^\vee)= {\cal P}(a^\vee)=a^\vee\cup a^\vee-a^\vee\cup_1 da^\vee$, which we integrate over $M_4^\vee=\mathbb{CP}^2$. In the $X$ basis a code state is a superposition of 2-cocycles $a^\vee$ of the dual cell complex, and the $Z$-checks shift $a^\vee$ by coboundaries. A hypercubic lattice is self-dual, so there the dual field again lives on a hypercubic lattice and the cup products of Sec.~\ref{sec:sgate} apply directly. On a triangulation the dual cells are not simplices; we treat that case below with a common simplicial refinement.
The logical gate is
\begin{equation}
    U(\omega^\vee,M_4^\vee)=i^{\int_{M_4^\vee} {\cal P}(a^\vee)}~.
\end{equation}
We write the explicit expression for the logical gate:
\begin{align}
    &U_\text{HSH}=U(\omega^\vee,M_4^\vee)\cr 
    &=
    \prod_{s_4^\vee\in M_4^\vee}\left(
    \prod_{s_2^\vee,s_2'^\vee\in s_4^\vee} i^{c_{s_4^\vee}(s_2^\vee,s_2'^\vee)\,a^\vee(s_2^\vee)a^\vee(s_2'^\vee)}\right)\cr
    &\cdot 
    \left(\prod_{\substack{s_2''^\vee,s_3^\vee\in s_4^\vee:\\ c^{(1)}_{s_4^\vee}(s_2''^\vee,s_3^\vee)\neq0}}\prod_{s_2'''^\vee\in\partial s_3^\vee}i^{o(s_2'''^\vee,s_3^\vee)a^\vee(s_2''^\vee)a^\vee(s_2'''^\vee)}\right)~,\label{eq:UHSH}
\end{align}
with the coefficients of Eq.~\eqref{eq:cupcoef} on the dual cells, and where the sign $o(s_2'''^\vee,s_3^\vee)=+ 1$ when $s_2'''^\vee$ appears in $\partial s_3^\vee$ with a $+$ sign, and $o(s_2'''^\vee,s_3^\vee)=- 1$ when $s_2'''^\vee$ appears in $\partial s_3^\vee$ with a $-$ sign.
In the above expression, for any 2-cell $s_2^\vee$ on the dual lattice, $a^\vee(s_2^\vee)=\frac{1-X_{s_2}}{2}$ for the 2-cell $s_2$ on the original lattice that intersects with $s_2^\vee$ at point. As for $U_\text{S}$, the sign of the correction term does not matter on code states, and we take the product as the definition of the circuit. On a hypercubic lattice $U_\text{HSH}$ has depth 2.

\textit{Dual cochain model on a triangulation.}
On a triangulation $K$ of $\mathbb{CP}^2$ the dual cells are polyhedra rather than simplices, so the cup and $\cup_1$ products in Eq.~\eqref{eq:UHSH} are not defined on them. We therefore move the dual field $a^\vee$ to a simplicial complex on which these products are defined, using a local map that preserves its cohomology class, and run the $S$ circuit~\eqref{eq:US} there. The construction has three steps.

\begin{enumerate}
\item \textit{A common refinement.} Let $K'$ be the barycentric subdivision of $K$. Its vertices are the barycenters $\hat\sigma$ of the simplices $\sigma$ of $K$, and its simplices are the flags $(\hat\sigma_0\cdots\hat\sigma_k)$ with $\sigma_0\subsetneq\dots\subsetneq\sigma_k$. Ordering the vertices of $K'$ by the dimension of $\sigma$ orders every simplex of $K'$ consistently, so the cup and $\cup_1$ products, and hence the circuit~\eqref{eq:US}, are defined on $K'$ without further choices. In particular, no vertex order on $K$ is needed. The refinement $K'$ also subdivides the dual complex $K^\vee$: the dual cell $D(\sigma)$ of a simplex $\sigma$ is the union of the simplices of $K'$ whose first vertex is $\hat\sigma$, and it has dimension $4-\dim\sigma$. The dual 2-cell $D(f)$ of a triangle $f$ carries $a^\vee(D(f))=(1-X_f)/2$. The closed dual block $\bar D(\sigma)=\bigcup_{\tau\supseteq\sigma}D(\tau)$ is a subcomplex of $K'$ and a cone with apex $\hat\sigma$, hence acyclic.

\item \textit{A local map from dual cochains to cochains on $K'$.} The dual blocks form a block complex that computes the homology of $K$ and of each closed block $\bar D(\sigma)$~\cite{Munkres:1984}. Assign to each simplex $(\hat\sigma_0\cdots\hat\sigma_k)$ of $K'$ the closed block $\bar D(\sigma_0)$ of its first vertex. Each face of the simplex starts with some $\hat\sigma_i$ with $\sigma_i\supseteq\sigma_0$, so its block lies inside $\bar D(\sigma_0)$, and every block is acyclic. The acyclic carrier theorem~\cite{Munkres:1984} therefore gives an augmentation-preserving chain map
\begin{equation}
\theta:\ C_*(K';\mathbb{Z}_2)\to C_*(K^\vee;\mathbb{Z}_2)
\end{equation}
that sends each simplex of $K'$ to a chain of dual cells in the closed block of its first vertex. The subdivision map $C_*(K^\vee)\to C_*(K')$ and $\theta$ are chain homotopy inverses, since both composites are carried by acyclic carriers that also carry the identity. The transpose $\theta^*$ therefore sends cocycles to cocycles and coboundaries to coboundaries, and it induces the canonical isomorphism $H^2(K^\vee;\mathbb{Z}_2)\cong H^2(K';\mathbb{Z}_2)\cong H^2(\mathbb{CP}^2;\mathbb{Z}_2)$. It is also local. For a triangle $s$ of $K'$ with first vertex $\hat\sigma_0$, $(\theta^*a^\vee)(s)$ is the parity of $a^\vee$ on the dual 2-cells in $\theta(s)$. These are dual to triangles of $K$ that contain $\sigma_0$, so $(\theta^*a^\vee)(s)$ is a parity of $X$-basis values of qubits on those triangles.

\item \textit{The circuit.} We set
\begin{equation}\label{eq:hshK}
U_\text{HSH}=i^{\int_{K'}{\cal P}(b)},\qquad b=\theta^*a^\vee~,
\end{equation}
that is, the circuit~\eqref{eq:US} on $K'$, with the orientation signs of $K'$ and with $a(s_2)$ replaced by the $0,1$ lift of $b(s_2)$ for every triangle $s_2$ of $K'$. On code states $a^\vee$ is a cocycle, so $b$ is a cocycle too, and the circuit evaluates $i^{\int{\cal P}(b)}$ as in Sec.~\ref{sec:sgate}. Since the evaluation of $a^{\vee}$ at each 2-simplex of $K'$ is expressed by the $X$-eigenvalues of original qubits, no auxiliary qubits are introduced.
\end{enumerate}

 \textit{Explicit form of $\theta$.}  We now construct $\theta$ on the vertices, edges, and triangles of $K'$, in that order. Since we only use how $\theta$ acts on the triangles, the following discussion suffices to complete the description of the logical operator.
 As mentioned above, the construction uses the fact that every closed dual block $\bar D(\sigma)$ is acyclic. All chains below have $\mathbb Z_2$ coefficients.

\begin{enumerate}
\item For each vertex $\hat\sigma$ of $K'$, choose a 4-simplex $p(\sigma)$ of $K$ containing $\sigma$, and define
\begin{align}
\theta(\hat\sigma)&=D(p(\sigma)).
\end{align}
Thus each vertex of $K'$ is assigned a dual vertex in $\bar D(\sigma)$.

\item Consider an edge $e=(\hat\sigma_0\hat\sigma_1)$, where $\sigma_0\subsetneq\sigma_1$. The images of its two endpoints have already been chosen, and both lie in $\bar D(\sigma_0)$. Since this block is connected, choose a path of dual edges within it joining these images. Define $\theta(e)$ to be the mod 2 sum of the dual edges along this path. By construction,
\begin{align}
\partial\theta(e)
&=\theta(\hat\sigma_0)+\theta(\hat\sigma_1)
=\theta(\partial e).
\end{align}
Each dual edge $D(t)$ along the path corresponds to a tetrahedron $t$ shared by two consecutive 4-simplices.

\item Consider a triangle $s=(\hat\sigma_0\hat\sigma_1\hat\sigma_2)$. Its three edges have already been assigned dual paths. Their sum
\begin{align}
z_s&=\theta(\partial s)
\end{align}
is a closed dual 1-chain contained in $\bar D(\sigma_0)$: the endpoint contributions cancel. Since
\begin{align}
H_1(\bar D(\sigma_0);\mathbb Z_2)&=0,
\end{align}
choose a dual 2-chain $c_s$ in this block satisfying $\partial c_s=z_s$, and define $\theta(s)=c_s$. Writing the chosen filling as a sum of dual 2-cells gives
\begin{align}
\theta(s)&=\sum_{f\in F(s)}D(f),\\
\partial\theta(s)&=\theta(\partial s),
\end{align}
where $F(s)$ is a set of triangles $f$ of $K$ containing $\sigma_0$.

The choices of vertices, paths, and fillings need not be unique. Acyclicity guarantees that the required paths and fillings exist. This constructs $\theta$ through degree two; repeating the same procedure in higher degrees gives the chain map in Eq. (23).

\end{enumerate}

Then $\theta(s)=\sum_{f\in F(s)}D(f)$ satisfies $\partial\theta(s)=\theta(\partial s)$. The sets $F(s)$ exist because $\bar D(\sigma_0)$ is acyclic, and the higher-degree part of $\theta$ exists by the same theorem but does not enter the circuit. 

\textit{The circuit on the original qubits.} With $X_F=\prod_{f\in F}X_f$, the value of $b$ on a triangle $s$ of $K'$ is the operator $\ell(s)=(1-X_{F(s)})/2$, the parity of the $X$-basis values on $F(s)$, and Eq.~\eqref{eq:hshK} becomes
\begin{align}
U_\text{HSH}&=\prod_{v}U_v~,\notag\\
U_v&=\prod_{s_4}\bigg(i^{\epsilon(s_4)\,\ell(\hat v\hat e\hat f)\,\ell(\hat f\hat t\hat p)}\notag\\
&\qquad\cdot\prod_{(s_2,s_3)}\ \prod_{s_2'\in\partial s_3}i^{o(s_2',s_3)\,\ell(s_2)\,\ell(s_2')}\bigg)~.\label{eq:hshexplicit}
\end{align}
Here $v$ runs over the vertices of $K$, and $s_4=(\hat v\hat e\hat f\hat t\hat p)$ over the 4-simplices of $K'$ with first vertex $v$, that is, over the full flags $v\subset e\subset f\subset t\subset p$ of $K$. The first factor is the cup product of the front and back triangles of $s_4$, and $(s_2,s_3)$ runs over the two $\cup_1$ pairs $\big((\hat v\hat t\hat p),(\hat v\hat e\hat f\hat t)\big)$ and $\big((\hat v\hat e\hat p),(\hat e\hat f\hat t\hat p)\big)$ of Sec.~\ref{sec:sgate}. In terms of Pauli operators, $i^{\ell(s)\ell(s')}=\exp[\frac{i\pi}{8}(1-X_{F(s)})(1-X_{F(s')})]$.

The circuit is local. Every triangle of $s_4$ starts with a simplex that contains $v$, so every set $F$ in $U_v$ consists of triangles of $K$ that contain $v$, and $U_v$ acts only on their qubits. $U_\text{HSH}$ is therefore a product of commuting unitaries $U_v$, diagonal in the $X$ basis, each on the qubits around one vertex of $K$. Each factor $i^{\pm\ell(s)\ell(s')}$ can be compiled, between two layers of Hadamard gates, from CNOT, $S$ and controlled-$S$ gates on those qubits. Unlike on a hypercubic lattice, the gates act on more than two qubits. Theorem~\ref{thm:HSH} shows that this circuit implements the logical $HSH$ gate.

\begin{theorem}\label{thm:HSH}
    Let $K$ be a triangulation of $M_4=\mathbb{CP}^2$, oriented as in Eq.~\eqref{eq:fundchain}, and define $U_\text{HSH}$ by the dual cochain model~\eqref{eq:hshK}. Then $U_\text{HSH}$ implements the logical HSH gate of the 2-form $\mathbb{CP}^2$ code. It is a constant-depth local unitary on any family of triangulations with bounded incidence.
\end{theorem}
\begin{proof}
The map $\theta^*$ moves the dual field to $K'$ without changing its cohomology class, so the circuit evaluates the same Pontryagin-square invariant as $U_\text{S}$, now on the $X$-basis logical labels. On a code state in the $X$ basis, $a^\vee$ is a 2-cocycle of $K^\vee$ whose class is $n^\vee\omega$, where $n^\vee=0,1$ labels the eigenstates $|\bar+\rangle$, $|\bar-\rangle$ of $\bar X$. Then $b=\theta^*a^\vee$ is a 2-cocycle of $K'$ in the class $n^\vee\omega$. Since $\int_{K'}{\cal P}(b)$ modulo 4 depends only on this class, it equals $(n^\vee)^2\int\omega^2=n^\vee$, so $U_\text{HSH}$ multiplies $|\bar+\rangle$ by $1$ and $|\bar-\rangle$ by $i$, which is the action of $HSH$.  With bounded incidence, boundedly many of the local gates $U_v$ act on each qubit, each on boundedly many qubits, so the greedy schedule of Sec.~\ref{sec:gates} gives bounded depth.
\end{proof}

\subsection{Constant-depth local unitary Hadamard gate in 2-form $\mathbb{CP}^2$ code}\label{sec:hgate}

The Hadamard logical gate is obtained by combining the $S$ and $HSH$ gates as 
    \begin{equation}
    \overline{S}\cdot \overline{HSH}\cdot \overline{S}=\frac{1+i}{\sqrt{2}}\overline{H}~.
\end{equation}
This identity can be checked directly from the $2\times2$ matrices of $S$ and $H$. Another way to understand it is from how the excitations are mapped under the symmetries: denote the excitation by $(q_e,q_m)$ for $q_e$ copies of electric excitation fused with $q_m$ copies of the magnetic excitation (where $q_e,q_m=0,1$ mod 2),
\begin{equation}
    (q_e,q_m)\rightarrow (q_e+q_m,q_m)\rightarrow (q_e+q_m,q_e)\rightarrow (q_m,q_e)~,
\end{equation}
which is the action of the Hadamard gate on the excitations.
Since the $S$ and $HSH$ circuits are constant-depth local unitaries, so is the resulting logical Hadamard gate, $U_\text{S}\,U_\text{HSH}\,U_\text{S}$ up to a global phase.

\subsection{Constant-depth local unitary CCZ gate in 2-form $\mathbb{CP}^3$ code}\label{sec:ccz}

Let us take three copies of the 2-form $\mathbb{CP}^3$ code. In other words, 
there are three types of physical qubits on each 2-cell in $M_6=\mathbb{CP}^3$ space. Denote their gauge fields by $a=(a_1,a_2,a_3)$.
Consider the cocycle $\omega(a)=a_1\cup a_2\cup a_3$. The corresponding logical gate is
\begin{equation}
    U(\omega,M_6)=(-1)^{\int_{M_6}a_1\cup a_2\cup a_3}~,
\end{equation}
which is a constant-depth local unitary (depth 3 on a hypercubic lattice). The explicit expression is
\begin{small}
\begin{align}
    &U_\text{CCZ}=U(\omega,M_6)\cr 
    &=\prod_{s_6\in M_6}\prod_{\substack{s_2,s_2',s_2''\in s_6:\\ c_{s_6}(s_2,s_2',s_2'')\neq0}} (-1)^{a_1(s_2)a_2(s_2')a_3(s_2'')},
\end{align}
\end{small}
where $a_i(s_2)=(1-Z_{s_2}^i)/2$ for any 2-cell $s_2$, and $Z^i$ is the physical Pauli operator for the $i^\text{th}$ type physical qubit.

\begin{theorem}\label{thm:CCZ}
    On a triangulation of $\mathbb{CP}^3$ with ordered vertices, the circuit $U_\text{CCZ}$
    implements the logical CCZ gate between three copies of the 2-form $\mathbb{CP}^3$ code. It is a constant-depth local unitary on any family of triangulations with bounded incidence, and on an explicit triangulation $\mathbb{CP}^3_{18}$ in the label order a complete schedule has depth 27 (Sec.~\ref{sec:cczprep}).
\end{theorem}
\begin{proof}
As for $U_\text{CZ}$, the phase depends only on the cohomology classes of $a_1$, $a_2$ and $a_3$, so it suffices to evaluate it on the logical labels. The ground states can be labeled by the flat connections \begin{equation}\label{eqn:logicalqubit}
    a_1=n_1 \omega,\quad a_2=n_2 \omega,\quad a_3=n_3 \omega~,
\end{equation}
 where $\omega$ is the K\"ahler form on $\mathbb{CP}^3$, and $n_i=0,1$ labels the ground states.
The circuit acts on the ground states $|n_1,n_2,n_3\rangle$ as 
    \begin{align*}
        &U_\text{CCZ}|n_1,n_2,n_3\rangle=(-1)^{\int_{\mathbb{CP}^3} a_1\cup a_2\cup a_3}|n_1,n_2,n_3\rangle\cr 
        &=(-1)^{n_1n_2n_3}|n_1,n_2,n_3\rangle=\overline{\text{CCZ}}_{n_1,n_2,n_3}|n_1,n_2,n_3\rangle~,
    \end{align*}
    which is the logical CCZ gate.
\end{proof}

\subsection{Universal quantum computation by code switching}\label{sec:universal}

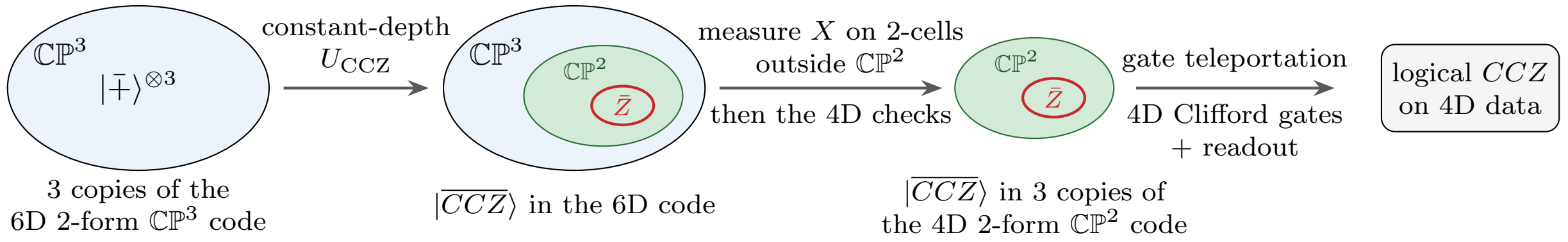
\begin{figure*}[t]
\centering
\begin{tikzpicture}[line join=round,
  arr/.style={-{Stealth[length=2.2mm]},thick,gray!70!black},
  lab/.style={font=\footnotesize,align=center}]
\begin{scope}
\draw[fill=zcol!8,draw=black] (0,0) ellipse (1.35 and 0.85);
\node at (0,0) {$|\bar+\rangle^{\otimes3}$};
\node[font=\small] at (-0.8,0.38) {$\mathbb{CP}^3$};
\node[lab] at (0,-1.2) {3 copies of the\\ 6D 2-form $\mathbb{CP}^3$ code};
\end{scope}
\draw[arr] (1.5,0)--(3.0,0);
\node[lab,above] at (2.25,0.05) {constant-depth\\ $U_\text{CCZ}$};
\begin{scope}[xshift=4.5cm]
\draw[fill=zcol!8,draw=black] (0,0) ellipse (1.35 and 0.85);
\draw[fill=cupB!22,draw=cupB!70!black] (0.3,-0.08) ellipse (0.82 and 0.52);
\draw[xcol,thick] (0.5,-0.18) ellipse (0.32 and 0.2);
\node[xcol,font=\scriptsize] at (0.5,-0.18) {$\bar Z$};
\node[font=\scriptsize,cupB!50!black] at (0.12,0.2) {$\mathbb{CP}^2$};
\node[font=\small] at (-0.8,0.38) {$\mathbb{CP}^3$};
\node[lab] at (0,-1.2) {$|\overline{CCZ}\rangle$ in the 6D code};
\end{scope}
\draw[arr] (6.0,0)--(8.3,0);
\node[lab,above] at (7.15,0.05) {measure $X$ on 2-cells\\ outside $\mathbb{CP}^2$};
\node[lab,below] at (7.15,-0.05) {then the 4D checks};
\begin{scope}[xshift=9.25cm]
\draw[fill=cupB!22,draw=cupB!70!black] (0,0) ellipse (0.82 and 0.52);
\draw[xcol,thick] (0.2,-0.1) ellipse (0.32 and 0.2);
\node[xcol,font=\scriptsize] at (0.2,-0.1) {$\bar Z$};
\node[font=\scriptsize,cupB!50!black] at (-0.18,0.28) {$\mathbb{CP}^2$};
\node[lab] at (0,-1.2) {$|\overline{CCZ}\rangle$ in 3 copies of\\ the 4D 2-form $\mathbb{CP}^2$ code};
\end{scope}
\draw[arr] (10.3,0)--(12.35,0);
\node[lab,above] at (11.32,0.05) {gate teleportation};
\node[lab,below] at (11.32,-0.05) {4D Clifford gates\\ + readout};
\node[draw,rounded corners,fill=gray!8,lab,minimum height=0.9cm] at (13.75,0) {logical $CCZ$\\ on 4D data};
\end{tikzpicture}
\caption{\label{fig:protocol}Universal computation by code switching (Sec.~\ref{sec:universal}). The constant-depth local unitary $CCZ$ gate of the 6D code prepares a magic state $|\overline{CCZ}\rangle$. Measuring Pauli $X$ on every 2-cell outside a $\mathbb{CP}^2$ subcomplex switches each copy to the 4D code. The logical $\bar Z$ is supported on a projective line $\mathbb{CP}^1$ inside that subcomplex (red ellipse, a schematic of this 2-sphere), so it commutes with these measurements, and the logical state is preserved up to a Pauli frame (Sec.~\ref{sec:universal}). Computation then runs entirely in the 4D code: its constant-depth Clifford gates and qubit-wise readout consume each magic state by gate teleportation, applying a logical $CCZ$ to the data.}
\end{figure*}

The 4D 2-form $\mathbb{CP}^2$ code has constant-depth logical $S$, $H$ and $CZ$ gates (Secs.~\ref{sec:czgate}--\ref{sec:hgate}), which generate the Clifford group. To make the gate set universal it suffices to supply $|CCZ\rangle$ magic states, which the 6D code produces with its constant-depth $CCZ$ gate. The protocol, summarized in Fig.~\ref{fig:protocol}, is:
\begin{enumerate}
\item Prepare three copies of the 2-form $\mathbb{CP}^3$ code in $|\bar+\rangle$, e.g.\ by initializing every physical qubit in $|+\rangle$ and measuring an independent generating set of the $Z$-stabilizers, with correction. (Measuring all of them would add redundancy, but Sec.~\ref{sec:cczprep} shows that the independent set gives a better state.)
\item Apply the constant-depth local unitary $U_\text{CCZ}$ to obtain $|\overline{CCZ}\rangle=\overline{CCZ}\,|\bar+\rangle^{\otimes3}$.
\item Switch each copy to the 4D code: measure Pauli $X$ on every 2-cell outside a $\mathbb{CP}^2$ subcomplex of the lattice (on an explicit triangulation $\mathbb{CP}^3_{18}$ of Sec.~\ref{sec:lattices}, the induced K\"uhnel $\mathbb{CP}^2_9$ on the nine vertices $b_1,\dots,b_9$), then apply the $Z$ correction and measure the stabilizers of the 2-form $\mathbb{CP}^2$ code on the remaining qubits, with error correction. In the absence of noise, this maps the logical state of each copy faithfully, up to a Pauli frame computed from the outcomes; the result is $|\overline{CCZ}\rangle$ in three copies of the 4D code.
\item Consume the magic state by the standard gate-teleportation circuit, which uses only Clifford gates and Pauli measurements in the 4D code, to apply a logical $CCZ$ gate. The corrections conditioned on the measurement outcomes are themselves Clifford ($CZ$ and Pauli gates), so the whole computation stays in the 4D code.
\end{enumerate}
Because both codes have the metacheck structure of Sec.~\ref{sec:singleshot}, we expect every step to need only
$O(1)$ rounds of error correction; a fault-tolerance analysis of the
whole protocol including the code switching, is left for future work.

The basis of the measurements in step~3 is important for getting the logical $CCZ$ state. Measuring $X$ removes
the qubits outside the subcomplex without reading out the gauge field
there: in the $Z$ basis a code state is a superposition of 2-cocycles
$a$, and the switch keeps its restriction to $\mathbb{CP}^2$, which
carries the same logical label $n$ because $\omega$ restricts to the
generator of $H^2(\mathbb{CP}^2;\mathbb{Z}_2)$. Measuring $Z$ instead
would read out $a$ outside the subcomplex, which can together read out the logical $\overline{Z}$ label and destroy the magic state.
Below we illustrate the steps of the code switching for the loop homological code on a generic closed manifold $M$ with a cellulation:
\begin{enumerate}
    \item Let $M'\subset M$ be a subcomplex of a cellulation $M$, and let $O$ be
the set of 2-cells of $M$ not in $M'$. Suppose that (i) no 2-cocycle of
$M$ in a nontrivial cohomology class is supported on $O$, and (ii) every
2-boundary of $M$ supported on $M'$ is a 2-boundary of $M'$. Then the
restriction $r\colon H^2(M;\mathbb{Z}_2)\to H^2(M';\mathbb{Z}_2)$ and
the inclusion $\iota\colon H_2(M';\mathbb{Z}_2)\to H_2(M;\mathbb{Z}_2)$
are isomorphisms, so the 2-form codes on $M$ and $M'$ encode the same
number of logical qubits. Starting from a code state of the code on $M$,
measure $X_f$ for every $f\in O$.
    \item Let $m$ be the 2-chain of the cells
with outcome $-1$, then choose a 2-chain $z$ of $M'$ with
$\partial' z=(\partial m)|_{M'}$, so that $m+z$ is a
2-cycle of $M$. Apply $Z(z)$ following the $X_f$ measurements.
After applying $Z(z)$, the qubits of $M'$ are in a code state of the code on
$M'$. The measurement outcomes carry no logical information, therefore the logical state
is unchanged up to a Pauli frame: for every 2-cycle $c$ of $M'$ the
logical operator $Z(c)$ is unchanged, and for every 2-cocycle $x$ of
$M$ the logical operator $X(x)$ becomes
$(-1)^{\langle x,\,m+z\rangle}X(x|_{M'})$.
\end{enumerate}

In the proposed protocol, we start from $M=\mathbb{CP}^3$ with a  $CCZ$ magic state and switch to the same logical state on $M'=\mathbb{CP}^2$.

\section{Practical Realization of 2-Form $\mathbb{CP}^3$ Code}
\label{sec:practical}

The previous sections used only the topology of $\mathbb{CP}^3$. In this section we turn the code into concrete stabilizer codes on explicit triangulations. We take a specific triangulation of $\mathbb{CP}^3$, read off the code from it using Table~\ref{tab:dictionary}, and compute its parameters exactly. We then build syndrome-extraction circuits and simulate them at the circuit level. All complexes were built and checked by computer (face counts, Euler characteristic, the pseudomanifold condition, and $\mathbb{Z}_2$ homology). All simulations use Stim~\cite{Gidney:2021stim} with belief-propagation decoding on the full circuit-level detector error model~\cite{Roffe:2020bposd}, with ordered-statistics post-processing for the $X$-basis memory of Sec.~\ref{sec:simulation} and the gap decoding of Sec.~\ref{sec:cczprep}.

\subsection{A $[[783,1,4]]$ code from the smallest known triangulation}
\label{sec:lattices}

\paragraph{The triangulation.} Any triangulation of $\mathbb{CP}^3$ needs at least 17 vertices~\cite{Arnoux:1991marin}. The smallest known one, $\mathbb{CP}^3_{18}$ of Bagchi and Datta~\cite{Bagchi:2012cp3}, has 18 vertices and 622 six-dimensional simplices (facets), listed explicitly in~\cite{Bagchi:2012cp3}. It is 2-neighborly, i.e., every pair of vertices spans an edge. Its numbers $f_i$ of $i$-simplices are listed in Table~\ref{tab:fvec}, together with their role in the code. As a check, the alternating sum $\sum_i(-1)^if_i=4$ equals the Euler characteristic of $\mathbb{CP}^3$.

\begin{table}[t]
\caption{\label{tab:fvec}Number $f_i$ of $i$-dimensional simplices in the triangulation $\mathbb{CP}^3_{18}$, and their role in the code (Table~\ref{tab:dictionary}).}
\begin{ruledtabular}
\begin{tabular}{rlr}
$i$ & role & $f_i$\\
\hline
0 & $X$-metachecks & 18\\
1 & $X$-checks & 153\\
2 & qubits & 783\\
3 & $Z$-checks & 2\,110\\
4 & $Z$-metachecks & 3\,021\\
5 & --- & 2\,177\\
6 & facets & 622\\
\hline
 & code & $[[783,1,4]]$\\
\end{tabular}
\end{ruledtabular}
\end{table}

\paragraph{The code.} With one qubit per triangle, the code has $n=f_2=783$ qubits, $153$ $X$-checks of weight 8 to 16, and $2\,110$ $Z$-checks of weight 4. The checks are highly redundant: only $136$ of the $X$-checks and $646$ of the $Z$-checks are independent. The number of logical qubits is therefore $k=783-136-646=1$, as required by $H_2(\mathbb{CP}^3;\mathbb{Z}_2)=\mathbb{Z}_2$.

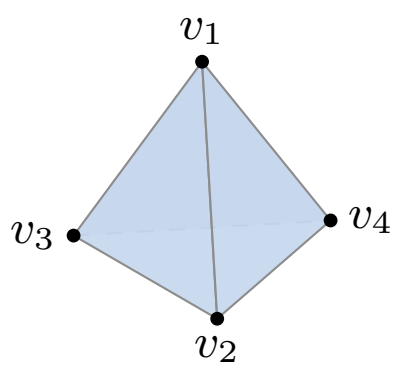
\begin{figure}[t]
\centering
\begin{tikzpicture}[scale=1.0,line join=round]
\coordinate (v1) at (0,1.45); \coordinate (v2) at (0.1,-0.25);
\coordinate (v3) at (-0.85,0.3); \coordinate (v4) at (0.85,0.4);
\fill[zcol!14] (v1)--(v3)--(v4)--cycle;
\draw[gray!70,dashed] (v3)--(v4);
\fill[zcol!26,fill opacity=0.9,draw=gray!90] (v1)--(v2)--(v3)--cycle;
\fill[zcol!26,fill opacity=0.9,draw=gray!90] (v1)--(v2)--(v4)--cycle;
\foreach \v in {v1,v2,v3,v4} \fill (\v) circle (1.3pt);
\node[above] at (v1) {\small$v_1$}; \node[below] at (v2) {\small$v_2$};
\node[left] at (v3) {\small$v_3$}; \node[right] at (v4) {\small$v_4$};
\end{tikzpicture}
\caption{\label{fig:emptytet}An empty tetrahedron. The four triangles of its boundary belong to the triangulation, but the solid tetrahedron $v_1v_2v_3v_4$ does not, so the product of $Z$ on these four triangles can be a nontrivial weight-4 logical. $\mathbb{CP}^3_{18}$ has 539 such logicals.}
\end{figure}

\begin{theorem}\label{thm:distance18}
The 2-form $\mathbb{CP}^3_{18}$ code is a $[[783,1,4]]$ code, with $d_Z=4$ and $91\le d_X\le 149$.
\end{theorem}
\begin{proof}
\textit{Idea.} A $Z$-type logical is a closed surface made of triangles. The smallest closed surface is the boundary of a tetrahedron, with four triangles, and it is a logical exactly when the solid tetrahedron is missing from the triangulation and its boundary is not a product of $Z$-checks. An $X$-type logical must cross every $Z$-type logical, so a large set of disjoint weight-4 $Z$-logicals bounds its weight from below.
A $Z$-type logical operator is a set $S$ of triangles forming a nontrivial 2-cycle: every edge lies in an even number of triangles of $S$.

\textit{No logical of weight below 4.} Each triangle has three edges, and each edge used by $S$ is shared by an even number of its triangles, so $3|S|=\sum_e(\text{number of triangles of $S$ at $e$})$ is even. Hence $|S|$ is even, and $|S|=2$ is impossible because two distinct triangles share at most one edge. So $|S|\ge4$. If $|S|=4$, the four triangles have six edges, each shared by exactly two of them, which forces them to be the four faces of a tetrahedron on four vertices $\{v_1,v_2,v_3,v_4\}$.

\textit{A logical of weight 4 exists.} If the solid tetrahedron $v_1v_2v_3v_4$ belongs to the triangulation, its boundary is a $Z$-check and hence trivial. If it does not, we call $v_1v_2v_3v_4$ an \textit{empty tetrahedron} (Fig.~\ref{fig:emptytet}), and its boundary may be a nontrivial cycle. $\mathbb{CP}^3_{18}$ has 545 empty tetrahedra. For 539 of them a rank computation shows that the boundary is not a sum of $Z$-checks, so each of these 539 is a weight-4 logical $\bar Z$, and $d_Z=4$.

\textit{Bounds on $d_X$.} An $X$-type logical must anticommute with every $\bar Z$, so it must contain an odd, hence nonzero, number of triangles from each of the 539 weight-4 logicals. We find 91 weight-4 logicals $\bar Z$ that are pairwise disjoint, i.e., share no triangle. Every $X$-logical contains a triangle of each of them and therefore has at least 91 triangles. An explicit $X$-logical of weight 149 gives the upper bound. 
Hence $d=\min(d_X,d_Z)=4$.
\end{proof}

The 9-vertex triangulation $\mathbb{CP}^2_9$ of K\"uhnel~\cite{Kuhnel:1983cp2} is embedded in $\mathbb{CP}^3_{18}$ as the subcomplex spanned by nine of the vertices, and all 36 empty tetrahedra of $\mathbb{CP}^2_9$ are among those of $\mathbb{CP}^3_{18}$.

The same computation on the 30-vertex triangulation $\mathbb{CP}^3_{30}$ of~\cite{Bagchi:2012cp3} gives a $[[2066,1,4]]$ code with 488 weight-4 logicals. 

\paragraph{Larger distance.}
Edgewise subdivisions~\cite{Edelsbrunner:2000egs} of $\mathbb{CP}^3_{30}$ give a family of codes with increasing distance, at the cost of more physical qubits. This refinement family also supports the asymptotic self-correction result of Theorem~\ref{thm:selfcorrection}. As a concrete example, the twofold subdivision yields a $[[103240,1,d]]$ code. For this code, we certify $d_Z\ge10$, while an explicit logical operator gives $d\le16$. The exact code distance remains undetermined.

\subsection{Syndrome-extraction circuits}
\label{sec:circuits}

We describe the syndrome-extraction circuit for the $[[783,1,4]]$ code, focusing on its depth and fault propagation. Each syndrome-extraction round consists of a $Z$ sub-round followed by an $X$ sub-round, with one bare ancilla per check.
The CNOT depth of a sub-round is at least the maximum degree of its Tanner graph connecting data qubits with ancilla qubits;
because Tanner graphs are bipartite, K\"onig's edge-colouring theorem
guarantees that this bound is achievable. 

For the $Z$ sub-round, the
bound is the maximum degree, given by the larger of the check weight $4$ and the maximal
number of checked tetrahedra containing a fixed triangle, which is $15$. For
the $X$ sub-round, it is the maximal number of triangles containing a fixed
edge, which is $16$. 
For sequential $Z$ and $X$ sub-rounds, our greedy schedule uses $15+17=32$ CNOT layers, one more than the lower bound of $15+16=31$.
We measure the full 
(overcomplete) set of checks, which provides the redundant metacheck
that would enable single-shot decoding (Sec.~\ref{sec:singleshot}).
The
resulting circuit has $3\,046$ qubits ($783$ data qubits and one ancilla
for each of the $153+2\,110$ checks), $10\,789$ CNOTs and $2\,263$
measurements per round, and its syndrome obeys the $3\,021$ weight-5
metachecks.

An ancilla fault can propagate through subsequent CNOTs to several data qubits, producing a correlated \textit{hook} error supported within the check being measured. Errors that differ by multiplication by a stabilizer have the same action on code states; we use this freedom when bounding the hook weights below.

A $Z$-check hook affects at most two faces of a filled tetrahedron, up to multiplication by that check. These two faces cannot belong to the same weight-4 logical $\bar Z$: the only weight-4 cycle containing both is the boundary of that tetrahedron, which is itself a $Z$-stabilizer. Thus a single fault cannot place two $Z$ errors on the support of any weight-4 logical. The $p^2$ scaling observed in the $X$-basis memory simulations of Sec.~\ref{sec:simulation}, whose circuits include hook errors, is consistent with this geometric protection. 

For $X$-type errors, a single fault affects at most eight triangles, modulo stabilizers. Since $d_X\ge91$, an undetected $X$-type logical error therefore requires at least $\lceil91/8\rceil=12$ faults. However, these support arguments do not determine the leading failure order of the implemented decoder. Different low-fault patterns can produce ambiguous syndromes, and the decoder can fail with fewer faults than the bound for an undetected logical error.

On lattices with larger $d_Z$, hook errors may overlap minimum-weight logicals more extensively. Choosing the CNOT order within each $Z$-check, or introducing flag qubits, may then be necessary to preserve the circuit-level distance.

\subsection{Circuit-level one-round benchmark}
\label{sec:simulation}

We benchmark the single-shot decoder by one-round preparation-and-readout
experiments with circuit-level CNOT and measurement noise: a logical
basis state is prepared, followed by one extraction round with
two-qubit depolarizing noise of strength $p$ after every CNOT,
independent bit flips of probability $p$ before every ancilla
measurement, and a noisy readout of every data qubit; initialization and
idling are noiseless.

Because the
two distances of the code are very different, $d_Z=4$ and $d_X\ge91$, we
test the two logical operators separately. In the $Z$ basis the data
start in $|0\rangle$ and the observable is a weight-4 logical $\bar Z$ on
the boundary of an essential empty tetrahedron; only errors with an $X$ component can flip this observable. After decoding, a logical flip corresponds to a residual nontrivial $X$-type logical operator, whose minimum weight is $d_X$.
In the $X$ basis the data start in
$|+\rangle$ and the observable is an $\bar X$ of weight $149$; only
$Z$-type faults can flip it, and they correspond to the $Z$ logicals with minimal weight 4.
Decoding is serial-schedule min-sum belief propagation on the exact
detector error model. For the $X$-basis benchmark, ordered-statistics decoding is used as a fallback when BP does not converge~\cite{Gidney:2021stim,Roffe:2020bposd}; the noiseless
circuits are verified to be deterministic. Appendix~\ref{app:sim}
describes how the logical error rates are obtained.

\begin{table}[t]
\caption{\label{tab:sim783}Measured one-round logical error rate of the
$[[783,1,4]]$ code (3\,046 qubits, 32 CNOT layers) in the $Z$ basis,
where only $X$-type faults can cause a logical error.}
\begin{ruledtabular}
\begin{tabular}{lll}
$p$ & fails/shots & $p_L^{(Z)}$ per round\\
\hline
$7\times10^{-2}$ & 60/100 & $0.60$\\
$5\times10^{-2}$ & 71/200 & $0.36$\\
$4\times10^{-2}$ & 27/200 & $0.14$\\
$3\times10^{-2}$ & 5/500 & $1.0\times10^{-2}$\\
$2\times10^{-2}$ & 0/700 & $<4.3\times10^{-3}$ (95\%)\\
$10^{-2}$ & 0/400 & $<7.5\times10^{-3}$ (95\%)\\
$\mathbf{10^{-3}}$ & \textbf{0/7\,000} & $\mathbf{<4.3\times10^{-4}}$ (95\%)\\
\end{tabular}
\end{ruledtabular}
\end{table}

\begin{table}[t]
\caption{\label{tab:simX}Measured one-round logical error rate of the
$[[783,1,4]]$ code in the $X$ basis, where only $Z$-type faults can
cause a logical error (BP+OSD; same circuit and noise as
Table~\ref{tab:sim783}).}
\begin{ruledtabular}
\begin{tabular}{lll}
$p$ & fails/shots & $p_L^{(X)}$ per round\\
\hline
$3\times10^{-3}$ & 943/2\,000 & $0.47$\\
$2\times10^{-3}$ & 619/2\,000 & $0.31$\\
$\mathbf{10^{-3}}$ & \textbf{930/12\,000} & $\mathbf{7.8\times10^{-2}}$\\
$5\times10^{-4}$ & 65/4\,000 & $1.6\times10^{-2}$\\
$3\times10^{-4}$ & 63/8\,000 & $7.9\times10^{-3}$\\
$2\times10^{-4}$ & 24/8\,000 & $3.0\times10^{-3}$\\
$10^{-4}$ & 21/24\,000 & $8.8\times10^{-4}$\\
$5\times10^{-5}$ & 10/40\,000 & $2.5\times10^{-4}$\\
\end{tabular}
\end{ruledtabular}
\end{table}

\begin{figure}[t]
\centering
\includegraphics[width=\columnwidth]{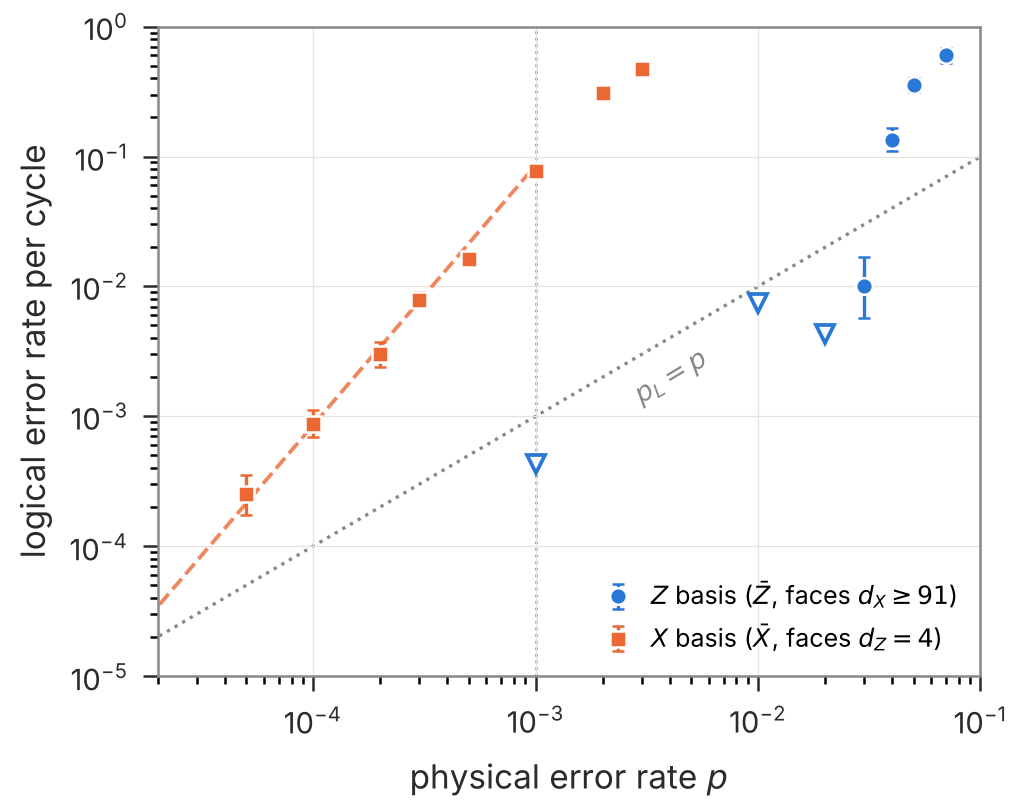}
\caption{\label{fig:errplot783}Logical error rate per round
for the $[[783,1,4]]$ code in the one-round benchmark with circuit-level
CNOT and measurement noise. Circles: the
logical $\bar Z$ (serial BP), which only $X$-type faults can flip;
open triangles are $95\%$ upper bounds at zero observed failures.
Squares: the logical $\bar X$ (BP+OSD), which only $Z$-type faults can
flip. Error bars are $68\%$ binomial (Clopper--Pearson) intervals. Dashed line: the fit
$p_L^{(X)}=8.6\times10^4\,p^2$ to the $X$-basis points with
$p\le3\times10^{-4}$. Dotted lines mark $p_L=p$ and $p=10^{-3}$.}
\end{figure}

\paragraph{$Z$ basis.} The results are collected in
Table~\ref{tab:sim783} and Fig.~\ref{fig:errplot783}. The logical error
rate drops sharply between $p=4\times10^{-2}$ and $3\times10^{-2}$ where it
falls below the physical error rate, so the pseudo-threshold of the $Z$-basis
memory lies between $3\%$ and $4\%$. At $p=10^{-3}$ we observe
{zero failures in 7\,000 decoded shots},
\begin{equation}
    p_L^{(Z)}\big(p=10^{-3}\big)\;<\;4.3\times10^{-4}\quad(95\%~\mathrm{CL}).
\end{equation}
Since an undetected $X$-type logical requires errors on at least $91$
triangles, which takes at least $12$ faults in this circuit
(Sec.~\ref{sec:circuits}), low-weight undetected fault patterns are
excluded. A BP decoder can still fail on fewer faults, so the leading
failure order of the decoded $Z$-basis memory is not established; we
expect, but have not shown, that its rate falls much faster than $p^2$
below the pseudo-threshold, and the bound at $p=10^{-3}$ is set by the
number of shots.

\paragraph{$X$ basis.} 
The $X$-basis memory behaves very differently
(Table~\ref{tab:simX} and Fig.~\ref{fig:errplot783}).
At $p=10^{-3}$, the decoded logical error probability per one-round trial is $p_L^{(X)}=7.8\times10^{-2}$, or approximately $7.8\%$. For $p\le3\times10^{-4}$
it follows $p_L^{(X)}\approx8.6\times10^{4}\,p^{2}$, so the logical
error rate would equal the physical one only near
$p\approx1.2\times10^{-5}$. 

In the $X$-basis benchmark, logical failures are caused by $Z$-type errors, for which the code has distance $d_Z=4$. Two errors on the support of a weight-4 logical $\bar Z$ produce the same syndrome as errors on the complementary two qubits. A decoder cannot distinguish these explanations from the syndrome alone. This ambiguity allows two-fault failures and provides a mechanism for the observed $p^2$ scaling, as discussed further in Sec.~\ref{sec:cczprep}.

A substantial contribution to the large prefactor comes from the $Z$-check measurements. Each triangle belongs to $10.8$ $Z$-checks on average, so its data qubit participates in about eleven CNOTs during the $Z$ sub-round. Faults in these gates, including errors propagated from the ancillas, increase the probability of $Z$ errors on the data. Removing the $Z$ sub-round reduces the logical error rate at $p=10^{-3}$ from $7.8\times10^{-2}$ to $6.8\times10^{-3}$ ($27/4\,000$), although this remains above the physical error rate $p$.

Thus, at $p=10^{-3}$, the $\bar Z$ memory meets a logical error target of $10^{-3}$, whereas the $\bar X$ memory does not. These results motivate lattices with larger $d_Z$ to suppress $Z$-type logical errors. Reducing the number of CNOTs per triangle could also lower the noise contribution, as observed for magic-state preparation in Sec.~\ref{sec:cczprep}. Measuring fewer redundant $Z$-checks, however, removes some of the consistency relations between check outcomes—the metachecks—that help the single-shot decoder identify measurement errors in the $Z$-basis benchmark.

\subsection{Magic-state preparation by correction and detection}
\label{sec:cczprep}

The non-Clifford layer of the universal gate set of Sec.~\ref{sec:gates} is the
magic state $|CCZ\rangle$, produced by preparing $|\bar+\rangle^{\otimes3}$ in
three copies of the 2-form $\mathbb{CP}^3$ code and applying the
circuit $U_{\rm CCZ}$. On $\mathbb{CP}^3_{18}$ this circuit evaluates the simplicial cup product
$\int a_1\cup a_2\cup a_3$ on each 6-simplex $(v_0\cdots v_6)$: it applies one physical CCZ gate to the front face
$(v_0v_1v_2)$ of copy 1, the middle face $(v_2v_3v_4)$ of copy 2, and the
back face $(v_4v_5v_6)$ of copy 3, so there are exactly $622$ physical CCZ gates (see Table \ref{tab:fvec}). Which triangle is the front, middle or back face depends on the vertex order that defines the cup product (Fig.~\ref{fig:cup}). We order the vertices by their labels in the facet list of Ref.~\cite{Bagchi:2012cp3}, $a_1<\cdots<a_9<b_1<\cdots<b_9$,
which we call the \textit{label order}; any other order gives an equally valid gate (Sec.~\ref{sec:proxy-noround}). We call the number of physical CCZ gates that act on the qubit of a triangle $t$ its \textit{leg count} $\ell_t$. The largest leg count bounds the depth of the circuit from below. In the label order the leg count is $0.79$ on average and at most $27$: the triangle $a_1a_2a_4$ is the front face of $27$ six-simplices, so the copy-1 qubit on it takes part in $27$ CCZ gates.

\paragraph{The preparation proxy.}
Each copy of the code is a code block that encodes one logical qubit, and each physical CCZ gate acts on three data qubits, one from each block. For the numerical benchmark we simulate only one block, with preparation and readout in the $X$ basis. We replace each physical CCZ gate by a single-qubit depolarizing channel acting on the participating qubit of the simulated block, so a qubit with leg count $\ell_t$ receives $\ell_t$ such channels. One noisy $X$-check round and a noisy qubit-wise $X$ readout follow (Appendix~\ref{app:ccz}). We report the logical $\bar X$ error probability among trials that pass the postselection criterion, together with the rejection rate, the fraction of trials that are discarded. This shows how much error correction and postselection lower the error rate, and how much of that reduction comes at the cost of discarding trials.

This single-block simulation is a heuristic benchmark of preparation noise, not a direct calculation or a bound on the infidelity of the accepted, encoded three-qubit $|CCZ\rangle$ state. Correlated errors between the three blocks are included only approximately (Appendix~\ref{app:ccz}), and neither code switching nor the subsequent use of the magic state is simulated.

We first describe the decoder (Sec.~\ref{sec:gapdecoding}) and apply it to the proxy without the measurement of the $Z$-checks that prepares $|\bar+\rangle$ (step~1 of Sec.~\ref{sec:universal}), to isolate the effect of the noise of the CCZ gates (Sec.~\ref{sec:proxy-noround}). We then include this measurement, for which measuring an independent set of the $Z$-checks turns out to be essential (Sec.~\ref{sec:proxy-zround}), and compare the result with magic-state
cultivation~\cite{Gidney:2024cultivation} (Sec.~\ref{sec:cultivation}).

\subsubsection{Correction and detection by gap decoding}
\label{sec:gapdecoding}

Paetznick
{et al.}~\cite{Paetznick:2026cad} demonstrated that a decoder which
corrects unambiguous syndromes and rejects ambiguous ones
raises the number of faults needed for an undetected logical error; for
a distance-4 code, from two to three. Their lookup-table decoder does not
scale to hundreds of qubits, so we implement a scalable equivalent,
two-class \textit{gap decoding}. Each syndrome is decoded twice by
BP+OSD~\cite{Roffe:2020bposd} on the detector error model, once
constrained to each value of the logical observable. Let $W_c$ be the
log-likelihood weight $\sum_i\log\frac{1-p_i}{p_i}$ of the solution in
class $c$. The decoder returns $\arg\min_c W_c$, and it rejects the shot
when the gap $|W_0-W_1|$ falls below a threshold $g$. The threshold
interpolates between plain correction ($g=0$) and detection-dominated
postselection (large $g$). 

\begin{figure}[t]
\centering
\begin{tikzpicture}[line join=round]
\foreach \k in {0,1,2} {
\begin{scope}[xshift={\k*2.75cm}]
\coordinate (A) at (0,0); \coordinate (B) at (1.7,0); \coordinate (C) at (0.85,1.472);
\coordinate (m1) at (0.85,0); \coordinate (m2) at (1.275,0.736); \coordinate (m3) at (0.425,0.736);
\draw[gray!80,fill=white] (A)--(B)--(C)--cycle;
\draw[gray!80] (m1)--(m2)--(m3)--cycle;
\end{scope}}
\fill[zcol!45,draw=gray!80] (0.85,0)--(1.275,0.736)--(0.425,0.736)--cycle;
\begin{scope}[xshift=2.75cm]
\fill[zcol!45,draw=gray!80] (0.85,0)--(1.275,0.736)--(0.425,0.736)--cycle;
\fill[zcol!45,draw=gray!80] (0,0)--(0.85,0)--(0.425,0.736)--cycle;
\end{scope}
\begin{scope}[xshift=5.5cm]
\fill[zcol!45,draw=gray!80] (0.85,0)--(1.275,0.736)--(0.425,0.736)--cycle;
\fill[zcol!45,draw=gray!80] (0,0)--(0.85,0)--(0.425,0.736)--cycle;
\fill[zcol!45,draw=gray!80] (0.85,0)--(1.7,0)--(1.275,0.736)--cycle;
\end{scope}
\node[align=center,font=\scriptsize] at (0.85,-0.62) {(a) true class: $\eta$\\ other class: $3\eta$\\ gap $2\eta$: corrected};
\node[align=center,font=\scriptsize] at (3.6,-0.62) {(b) true class: $2\eta$\\ other class: $2\eta$\\ gap $0$: rejected};
\node[align=center,font=\scriptsize] at (6.35,-0.62) {(c) true class: $3\eta$\\ other class: $\eta$\\ gap $2\eta$: wrong class};
\end{tikzpicture}
\caption{\label{fig:gaptoy}Gap decoding on a single weight-4 logical $L$,
whose four triangles are drawn as the unfolded net of the tetrahedron
boundary of Fig.~\ref{fig:emptytet}. Shaded triangles carry a fault,
each costing weight $\eta\approx7$ at $p=10^{-3}$. The syndrome is explained
either by the faults themselves (true class) or by their complement in
$L$ (other class, which differs by the logical operator on $L$).
(a)~One fault: the true class is cheaper by $2\eta$, and the fault is
corrected. (b)~Two faults: the classes tie, and the shot is rejected.
(c)~Three faults: the other class is cheaper by $2\eta$, and the decoder
confidently applies the wrong correction. A threshold $0<g<2\eta$ rejects
(b) and accepts (a) and (c), so an accepted logical error needs three
faults instead of two.}
\end{figure}
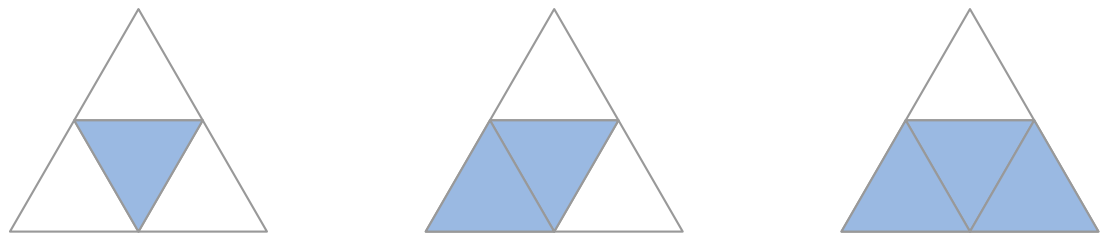

The mechanism is easiest to see on a single weight-4 logical $L$
(Fig.~\ref{fig:gaptoy}). At $p=10^{-3}$ each fault costs a weight
$\eta=\log\frac{1-p}{p}\approx7$. If one of the four qubits of $L$ is
flipped, the class of the actual error has weight $\eta$, while the other
class must flip the remaining three qubits and has weight $3\eta$; the gap
is $2\eta\approx14$ and the fault is corrected. If two are flipped, the
syndrome is explained equally well by those two faults and by the
complementary two, the classes tie at $2\eta$, and the shot is rejected
instead of being miscorrected half of the time. If three are flipped,
the syndrome looks like a single fault on the fourth qubit, the gap is
again $2\eta$ but now in favor of the wrong class, and the decoder
confidently applies the wrong correction. A threshold $0<g<2\eta$ therefore
removes the two-fault failures but not the three-fault ones, and an
accepted logical error needs three faults on one logical rather than
two. In a real shot, faults elsewhere and the spread of mechanism weights
blur these values. A typical shot at $p=10^{-3}$ contains a few isolated
faults, and its gap is set by the cheapest of them to move to the other
class, near $2\eta$; this is why the rejection in Table~\ref{tab:gap783}
rises steeply once $g$ approaches $2\eta$, from $10\%$ at $g=10$ to $47\%$
at $g=15$.
Since $\eta$ depends on $p$, a fixed threshold $g$ is a different fraction of
the single-fault gap $2\eta$ at different $p$.

\subsubsection{The proxy without the $Z$-check round}
\label{sec:proxy-noround}

\begin{table}[t]
\caption{\label{tab:gap783}Measured gap decoding of the
$|CCZ\rangle$-preparation proxy on the $[[783,1,4]]$ code at
$p=10^{-3}$ ($80\,000$ shots, one block: state preparation, all
$622$ CCZ legs, one noisy $X$-check round, noisy qubit-wise readout;
no $Z$-check round).
The last line, whose $95\%$ confidence interval is
$(0.7$--$4.3)\times10^{-4}$, agrees within a factor of two with the
three-fault estimate $w_3=1.1\times10^{-4}$ per prepared block
(Sec.~\ref{sec:proxy-noround}), i.e.\ $2.8\times10^{-4}$ per accepted block.}
\begin{ruledtabular}
\begin{tabular}{llll}
policy & rejection & fails/kept & $p_L\,|\,$accept\\
\hline
correction only & $0\%$ & 672/80\,000 & $8.4\times10^{-3}$\\
gap $\ge2$ & $1.6\%$ & 115/78\,688 & $1.5\times10^{-3}$\\
gap $\ge10$ & $9.9\%$ & 35/72\,050 & $4.9\times10^{-4}$\\
gap $\ge15$ & $47\%$ & 14/42\,023 & $3.3\times10^{-4}$\\
gap $\ge20$ & $62\%$ & 6/30\,463 & $2.0\times10^{-4}$\\
\end{tabular}
\end{ruledtabular}
\end{table}

\begin{figure*}[t]
\centering
\includegraphics[width=0.92\textwidth]{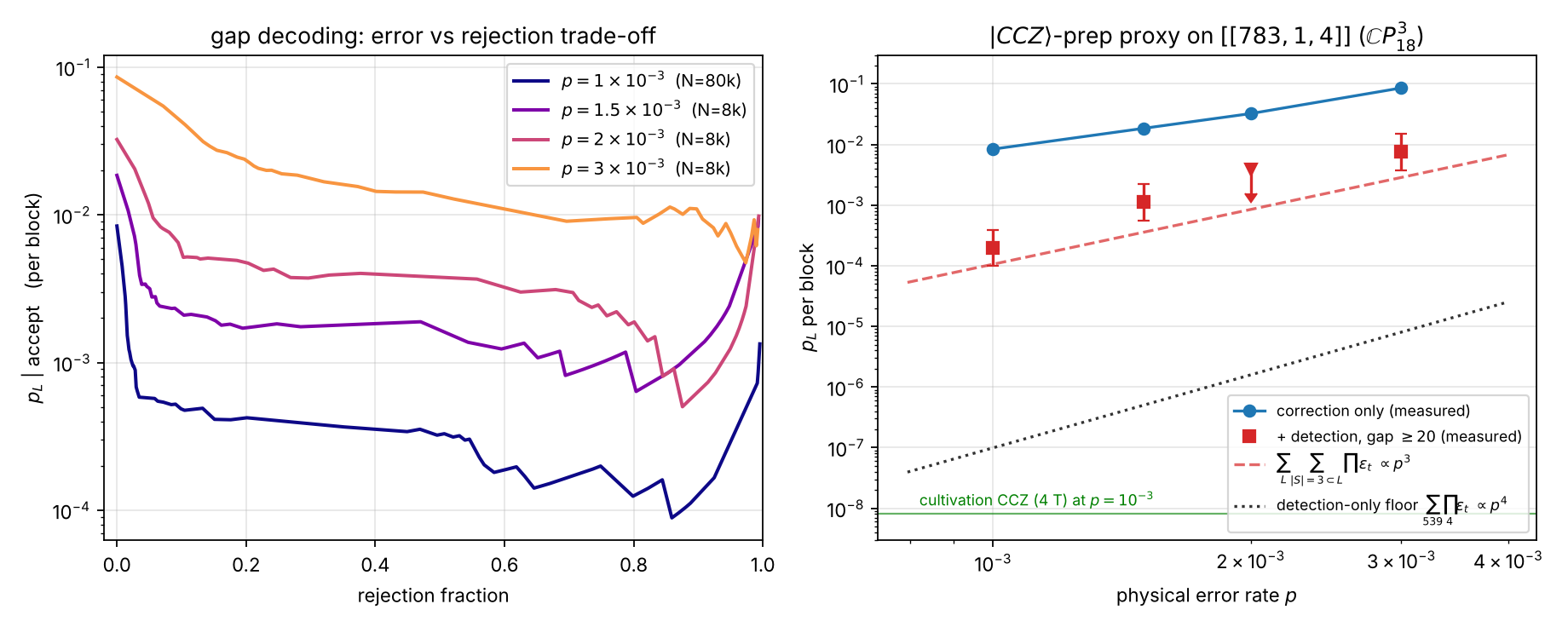}
\caption{\label{fig:gapdecoding}Correction and detection for the
$|CCZ\rangle$-preparation proxy on the $[[783,1,4]]$ code, without the
$Z$-check round. Left:
conditional logical error rate versus rejection fraction as the gap
threshold is swept, for $p=1$--$3\times10^{-3}$; the large markers
are the threshold $g\ge10$, with $68\%$ binomial intervals. Right:
measured correction-only and gap-decoded ($g\ge10$) error rates per
block, with $68\%$ binomial intervals; thin solid lines are
weighted least-squares power-law fits, with the exponents given in the
legend. Dashed: the leading-order three-fault estimate $w_3$ per
prepared block; dotted: the four-fault estimate $w_4$ for detection
only; the horizontal line marks a cultivation-based CCZ at $p=10^{-3}$.
Both panels are generated by one script from the per-shot data.}
\end{figure*}

Table~%
\ref{tab:gap783} and Fig.~\ref{fig:gapdecoding} give the result. The
noiseless circuit is deterministic, and its detector error model has no
undetectable single mechanism. At the operating point $p=10^{-3}$, gap
decoding lowers the accepted error rate by a factor of about $40$ at
$62\%$ rejection. 
It also
steepens the scaling from $p^2$ to $p^3$: over $p=(1$--$3)\times10^{-3}$
the correction-only rate scales as $p^{2.11\pm0.05}$, while at the fixed
threshold $g=10$ the accepted rate scales as $p^{3.1\pm0.2}$. 
This is consistent with two-to-three-fault shift
discussed in Ref.~\cite{Paetznick:2026cad}.

\subsubsection{Including the $Z$-check round}
\label{sec:proxy-zround}

\begin{figure}[t]
\centering
\includegraphics[width=\columnwidth]{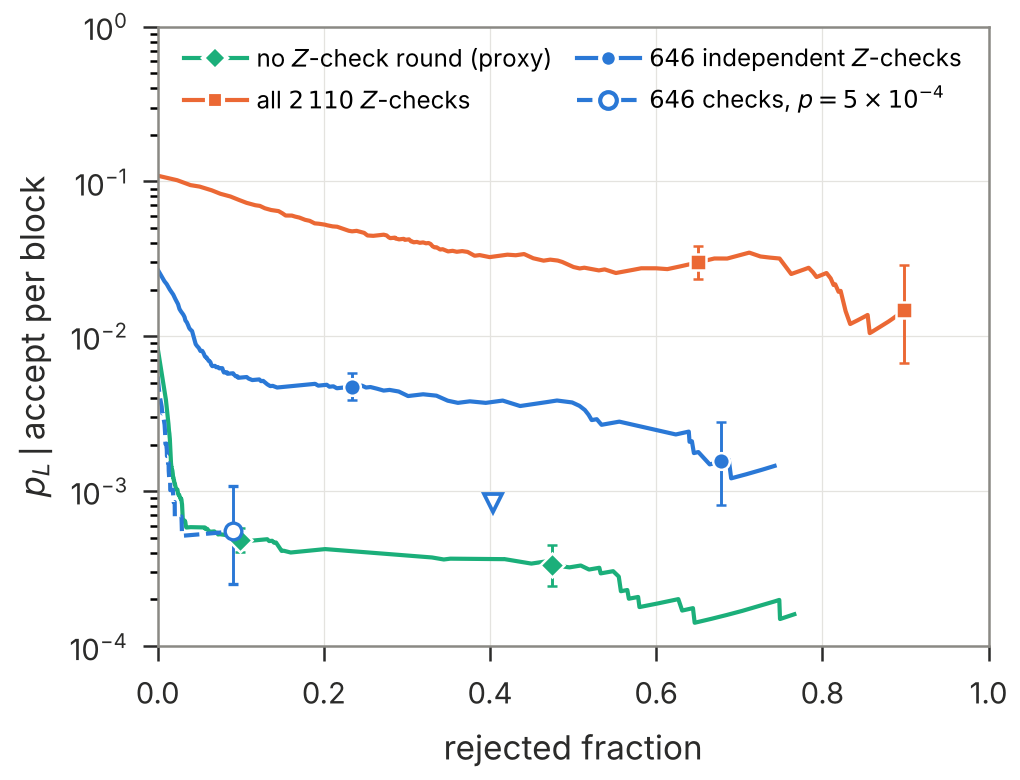}
\caption{\label{fig:zround}Accepted error of the $|CCZ\rangle$-preparation
proxy, with and without the $Z$-check measurement, on the $[[783,1,4]]$
code against the rejected fraction as
the gap threshold is swept, for the three ways of handling the $Z$-check
round of Table~\ref{tab:gapZ}: no round (the proxy of
Table~\ref{tab:gap783}), all $2\,110$ checks, and an independent set of
$646$, all at $p=10^{-3}$, and the independent set at
$p=5\times10^{-4}$. Markers are the thresholds $g\ge10$ and $g\ge15$,
with $68\%$ binomial intervals; the open triangle is the $95\%$ upper
bound at zero failures. Curves end where fewer than three accepted
failures remain.}
\end{figure}

Preparing $|\bar+\rangle$ requires measuring the $Z$-checks (step~1 of
Sec.~\ref{sec:universal}), which the proxy above omits. We add this
measurement between the preparation and the CCZ legs, with the same
noise, and compare two versions of it (Table~\ref{tab:gapZ} and
Fig.~\ref{fig:zround}). The first measures all $2\,110$ $Z$-checks, as in
the memory round of Sec.~\ref{sec:simulation}. The second measures an
independent generating set of $646$ checks, the rank of the $Z$-check
matrix. It prepares the same state, since every other check is a product
of measured ones. 
Both versions include the $Z$
errors that residual $X$ errors on the other two blocks place on the
simulated block through the CCZ gate (Appendix~\ref{app:ccz}). The result
is still a single-block proxy: the other blocks enter only through this
kickback model.

\begin{table*}[t]
\caption{\label{tab:gapZ}Gap decoding of the $|CCZ\rangle$-preparation
proxy including the $Z$-check measurement that prepares $|\bar+\rangle$,
on the $[[783,1,4]]$ code (one block, $X$ basis;
Appendix~\ref{app:ccz}). The round measures either all $2\,110$
$Z$-checks or an independent generating set of $646$; both include, as
a model, the $Z$ errors kicked back through the CCZ gate by residual $X$
errors on the other blocks. The first line repeats the
proxy without the round (Table~\ref{tab:gap783}) for reference. With no
failure we quote the one-sided $95\%$ bound $3/N$.}
\begin{ruledtabular}
\begin{tabular}{llrllllllll}
 & & & correction only & \multicolumn{3}{c}{gap $\ge10$} & \multicolumn{3}{c}{gap $\ge15$}\\
$Z$-check round & $p$ & shots & $p_L$ & rejection & fails/kept & $p_L\,|\,$accept & rejection & fails/kept & $p_L\,|\,$accept\\
\hline
none (proxy) & $10^{-3}$ & $80\,000$ & $8.4\times10^{-3}$ & $9.9\%$ & 35/72\,050 & $4.9\times10^{-4}$ & $47\%$ & 14/42\,023 & $3.3\times10^{-4}$\\
all $2\,110$ & $10^{-3}$ & $2\,000$ & $1.1\times10^{-1}$ & $65\%$ & 21/699 & $3.0\times10^{-2}$ & $90\%$ & 3/204 & $1.5\times10^{-2}$\\
$646$ independent & $2\times10^{-3}$ & $2\,000$ & $1.2\times10^{-1}$ & $63\%$ & 27/738 & $3.7\times10^{-2}$ & $92\%$ & 4/169 & $2.4\times10^{-2}$\\
 & $\mathbf{10^{-3}}$ & $8\,000$ & $2.7\times10^{-2}$ & $\mathbf{23\%}$ & \textbf{29/6\,125} & $\mathbf{4.7\times10^{-3}}$ & $68\%$ & 4/2\,572 & $1.6\times10^{-3}$\\
 & $5\times10^{-4}$ & $6\,000$ & $5.5\times10^{-3}$ & $9.1\%$ & 3/5\,457 & $5.5\times10^{-4}$ & $40\%$ & 0/3\,581 & $<8.4\times10^{-4}$\\
\end{tabular}
\end{ruledtabular}
\end{table*}

Each triangle lies in $10.8$
$Z$-checks on average, so measuring all of them places about eleven
further CNOTs, and their $Z$-type faults, on every data qubit. This
raises the marginal $Z$-error rate of a triangle from about $3.8p$ in
the proxy to about $10p$. A shot at $p=10^{-3}$ then contains about ten
faults, and failures typically involve four or five faults spread over
several weight-4 logicals, which a gap
threshold cannot isolate: the error of $1.1\times10^{-1}$ with
correction only falls just to $3.0\times10^{-2}$ at $65\%$ rejection.
Meanwhile, measuring the independent set of $Z$-checks places three or four CNOTs on most triangles, for a
total rate of about $6p$ and about six faults per shot.
Gap decoding then gives
$4.7\times10^{-3}$ at $23\%$ rejection with $g\ge10$, and
$1.6\times10^{-3}$ at $68\%$ rejection with $g\ge15$.

The error falls steeply with $p$
(Table~\ref{tab:gapZ} and Fig.~\ref{fig:zround}). At $p=5\times10^{-4}$
measuring the independent set of $Z$-checks gives $5.5\times10^{-4}$ at $9\%$ rejection with
$g\ge10$ ($3$ failures in $5\,457$ accepted shots), and no failure in
$3\,581$ accepted shots at $40\%$ rejection with $g\ge15$. From
$p=5\times10^{-4}$ to $10^{-3}$ the accepted error at $g\ge10$ grows
about 9 times, close to the $p^3$ scaling of the errors.

\subsubsection{Comparison with magic-state cultivation}
\label{sec:cultivation}

Magic-state cultivation~\cite{Gidney:2024cultivation} currently reaches
$2\times10^{-9}$ per $T$ state at $p=10^{-3}$. A $|CCZ\rangle$ assembled
from four cultivated $T$ states~\cite{Jones:2013toffoli} therefore has
an error of $\approx8\times10^{-9}$, before the errors of the
distance-$15$ surface-code Clifford circuit that converts them into a
$|CCZ\rangle$ state. The accepted error of the $[[783,1,4]]$ preparation
with the $Z$-check round, $4.7\times10^{-3}$ per block at $23\%$
rejection and $1.6\times10^{-3}$ at $68\%$ at $p=10^{-3}$, and
$5.5\times10^{-4}$ at $9\%$ at $p=5\times10^{-4}$, remains far above this
benchmark (Fig.~\ref{fig:gapdecoding}), as expected for a distance-4
code in which three faults suffice for an accepted error.

The cup-product construction prepares an encoded $|CCZ\rangle$ directly in the 6D code, without assembling it from $T$ states. After code switching (Sec.~\ref{sec:universal}), the state can be consumed using the constant-depth Clifford gates of the 4D code (Sec.~\ref{sec:gates}).
To narrow the performance gap with cultivation, the main priority is to increase $d_Z$. The current value, $d_Z=4$, leaves the preparation vulnerable to phase errors and limits the quality of the accepted states.

\subsection{Hardware feasibility}
\label{sec:hardware}

We close the practical analysis by placing the $[[783,1,4]]$ code
against demonstrated hardware, in terms of qubit number, connectivity, and the measurement
weights.

\paragraph{Qubit budgets.} A full $[[783,1,4]]$ block --- data, all ancillas,
one noisy round per logical cycle --- is $3\,046$ physical qubits with
$32$ CNOT layers and $2\,263$ measurements per round. The three-block
magic-state preparation of Sec.~\ref{sec:cczprep}, which measures an
independent set of $646$ $Z$-checks per block, needs
$3\times(783+646+153)=4\,746$ qubits. Both fit within the $6\,100$ atoms already trapped in a single
neutral-atom tweezer array~\cite{Manetsch:2024tweezer}, and logical-qubit
processors with hundreds of atoms and mid-circuit readout have been
demonstrated~\cite{Bluvstein:2023logical}. The operating point
$p=10^{-3}$ assumed throughout equals the two-qubit error rate of
trapped-ion QCCD processors~\cite{Moses:2023racetrack}, the platform on
which correction-and-detection decoding was
demonstrated~\cite{Paetznick:2026cad}. We note that these capabilities come from
different devices and platforms: the atom count of
Ref.~\cite{Manetsch:2024tweezer} is for trapping and imaging, and no
current device combines this size with the error rate $p=10^{-3}$.

\paragraph{Connectivity.} A six-dimensional code has no local embedding in
two or three spatial dimensions. A planar architecture with fixed
nearest-neighbour couplings could still run it by routing, at a cost in
depth and qubits that grows with the system size, but it would lose the
constant-depth local implementation assumed here. We instead assume the
reconfigurable,
effectively all-to-all connectivity provided by atom transport and ion
shuttling~\cite{Bluvstein:2023logical,Moses:2023racetrack}.
The graph relevant to the $\mathbb{CP}^3$ code is sparse and of bounded degree: each data
qubit couples to its $3$ edge ($X$) checks and on average $10.8$
tetrahedron ($Z$) checks in the $[[783,1,4]]$ code, and each syndrome
round is a fixed sequence of $32$ disjoint two-qubit layers, a pattern
suited to the parallel, batched atom moves of tweezer arrays.

\paragraph{Weights and native gates.} The $Z$-stabilizers have fixed weight
$4$ (weight $6$ on hypercubic lattices); only the $X$-stabilizers are
heavy (weight $8$--$16$ on $\mathbb{CP}^3_{18}$), and our circuits
measure them directly. The
non-Clifford layer is the $622$ physical CCZ gates of
Sec.~\ref{sec:cczprep}, applicable either as native multiply-controlled
Rydberg gates or as constant-depth two-qubit decompositions.

\section{Comparison with Universal Computation in the 6D Color Code}
\label{sec:colorcode}

Ref.~\cite{Bombin_2013}
constructed the first self-correcting quantum computer from topological
stabilizer codes based on 6D color codes. Since
that proposal is the closest relative of the present one, we compare
the two architectures point by point.

\paragraph{Completing the gate set.} The 6D color code supports transversal
CNOT and $R_2$, but does not support a transversal Hadamard gate.
In our scheme the entire single-qubit Clifford
group is implemented by constant-depth local unitaries of Sec.~\ref{sec:gates},
while the constant-depth local circuit CCZ gate of the 6D code supplies the
non-Clifford resource. 

\paragraph{Stabilizer weights.} The stabilizer weights of a color code
depend on the colex. As a reference we take a colex whose $n$-cells are
$n$-cubes, with $2^n$ vertices each; its 6D color-code stabilizers reach
weight $2^6=64$ (weight $8$ on the smaller cells). The 2-form $\mathbb{CP}^3$ code has $w_Z=6$ on the
hypercubic lattice ($w_Z=4$ on triangulations) with only $w_X$ growing;
Sec.~\ref{sec:practical} realizes an instance with $w_Z=4$ and $w_X\le16$
explicitly.

\paragraph{Passive and active memory.} Ref.~\cite{Bombin_2013} operates
as a finite-temperature passive memory: initialization uses controllable
Hamiltonian couplings or external fields, and transversal gates must be
applied fast on the scale of the protecting Hamiltonian. The 2-form $\mathbb{CP}^3$ code is also a
passive memory: by Theorem~\ref{thm:selfcorrection}, its Hamiltonian
protects the logical qubit at nonzero temperature on refined lattices.
Our protocol, however, uses it as an active stabilizer-circuit code. The
same extended-excitation structure that yields self-correction manifests
as the metacheck structure that motivates single-shot
decoding~\cite{Bombin:2014ksv}, so protection is expected from one round
of noisy syndrome measurement per logical cycle with metacheck repair
(Sec.~\ref{sec:singleshot}), with no Hamiltonian engineering assumed.

\section{Discussion and Outlook}
\label{sec:discussion}

We have developed the 2-form $\mathbb{CP}^3$ code from its topological
definition to practical circuits. On families of refined lattices it is a
provably self-correcting quantum memory for a single logical qubit. It
has a $CCZ$ gate implemented by a constant-depth local circuit. 
The 4D
2-form $\mathbb{CP}^2$ code has all Clifford gates with constant depth unitary circuits, and together with
code switching between $\mathbb{CP}^2$ and $\mathbb{CP}^3$ code, they provide a proposed protocol for universal
computation. Fault tolerance of this protocol, expected to need $O(1)$ rounds per
logical operation due to the metacheck structure, remains to be established. 

Using the smallest known triangulation of $\mathbb{CP}^3$, we obtain a $[[783,1,4]]$ code and construct explicit circuits for syndrome extraction and the logical CCZ gate. Circuit-level simulations provide a proof of concept for one-round memory and a single-block proxy for magic-state preparation. The memory benchmarks show substantially stronger protection of $\bar Z$ than of $\bar X$, while the preparation proxy demonstrates that combining error correction with postselection can reduce the accepted error rate at the cost of a lower acceptance probability.

The main practical limitation is the small distance $d_Z=4$, which leaves both the $\bar X$ memory and magic-state preparation vulnerable to phase errors. Improving $d_Z$ while controlling circuit noise and qubit overhead is therefore a central goal. Establishing thresholds, evaluating larger lattices, and testing repeated recovery and the full computational protocol remain tasks for future work.

\section*{Acknowledgment}

P.-S.H. is supported by Department of Mathematics King’s College
London. 
R.K. is supported by JST PRESTO Grant Number \texttt{JPMJPR26A8}, Japan, JSPS KAKENHI Grant-in-Aid \texttt{No.26K24517}, and the Department of Applied Physics, the University of Tokyo.   
Z.W. is partially supported by \texttt{ARO W911NF261A263}.  Claude Opus 5.5 and GPT-6.1 Sol were used to assist with drafting and editing text and with writing and running the scripts for the computations. The authors are responsible for the content of the paper.
All scripts and data used for the computations in this paper are available from the authors upon request.

\appendix

\section{Proof of self-correction on refined lattices}
\label{app:selfcorrection}

This appendix proves Theorem~\ref{thm:selfcorrection}. We first explain the idea and the structure of the proof, then give the details.

\paragraph{Idea of the proof.} Thermal noise acts on the memory through single-qubit Pauli errors, and the stored state is lost only when the accumulated error becomes a logical operator that the decoder cannot undo. In the 2-form code every syndrome is a union of closed objects: loops of violated edges for $Z$-type errors, and closed $(2N-3)$-dimensional objects on the dual lattice for $X$-type errors. Each violated check costs energy $2J$, so a syndrome component costs energy proportional to its size. A small component can be removed locally, by filling it with a membrane inside a small contractible region, and such a local filling does not change the logical class. To change the logical class, the error must therefore pass through a configuration with a large syndrome component; this is the energy barrier. Below a critical temperature, such configurations are exponentially rare in thermal equilibrium, because the Boltzmann factor of a large component outweighs the number of components of that size (a Peierls argument). The thermal dynamics has to cross this rare set to change the logical class, so the memory time is exponentially long in the size of the barrier.

\paragraph{Structure of the proof.} The argument follows the proof for the 4D toric code~\cite{doi:10.1142/S1230161210000023} and makes these steps precise:
\begin{enumerate}
\item \textit{The lattice} (Sec.~\ref{app:sc-lattice}). We need two properties of the family $K_m$ that hold uniformly in $m$: bounded local geometry, which controls the counting in the Peierls argument, and a coarser triangulation $K_q$ whose vertex stars are contractible regions of linear size $m/q$ in units of $K_m$, inside which small syndrome components can be filled.
\item \textit{The decoder and the energy barrier} (Sec.~\ref{app:sc-decoder}). A local-filling decoder fills every small syndrome component inside one such star and thereby assigns a logical class to every error whose syndrome has no large component. A single-qubit change of the error can change this class only when a large component is present.
\item \textit{The Peierls bound} (Sec.~\ref{app:sc-peierls}). Large components are exponentially rare in the Gibbs distribution. This step uses $H_1(\mathbb{CP}^N)=H^3(\mathbb{CP}^N)=0$, which makes every closed loop, and every closed object of the dual kind, the syndrome of some error.
\item \textit{The dynamics} (Secs.~\ref{app:sc-davies} and~\ref{app:sc-proof}). Davies dynamics with single-qubit couplings acts on code states dressed by Pauli errors as a classical Markov chain on the errors, reversible with respect to the Gibbs distribution. A reversible chain whose logical class can change only through an exponentially rare set needs an exponentially long time to change it.
\end{enumerate}
The lattice enters only through the two properties of step~1. Ref.~\cite{Bombin_2013} gives a sufficient condition of a similar kind, based on the connectedness of excitations, for topological stabilizer codes on regular lattices. 

Throughout, we assume that $N\ge2$, and homology and cohomology in the proof have $\mathbb{Z}_2$ coefficients. 
A $Z$-type Pauli error is a 2-chain $E_Z$; it violates the $X$-checks on the edges of $\partial E_Z$. An $X$-type error is a 2-cochain $E_X$; it violates the $Z$-checks on the 3-simplices of $\delta E_X$. We write $E=(E_X,E_Z)$ and $P_E=X(E_X)Z(E_Z)$, where $X(c)=\prod_{f\in c}X_f$ and $Z(c)$ is defined in the same way. Measured from the ground-state energy, the energy of $P_E$ applied to a code state is
\begin{equation}\label{eq:sc-energy}
\varepsilon(E)=2J\big(|\delta E_X|+|\partial E_Z|\big)~.
\end{equation}

\subsection{Edgewise subdivisions}
\label{app:sc-lattice}

This subsection establishes the two lattice properties used in the rest of the proof. Lemma~\ref{lem:sc-degree} bounds the local geometry of $K_m$ uniformly in $m$. Lemmas~\ref{lem:sc-coords} and~\ref{lem:sc-acyclic} compare $K_m$ with a coarser subdivision $K_q$: a simplex of $K_m$ is small on the scale of $K_q$, and the star of a vertex of $K_q$, or of $K$, is a contractible region in which every cycle bounds.

Let $K$ be a triangulation of $\mathbb{CP}^N$ with totally ordered vertices. We regard each simplex $\sigma=[v_0<\cdots<v_k]$ as the convex hull of its vertices. A point $x\in\sigma$ is uniquely specified by its barycentric coordinates, the real weights $t_{v_i}(x)$ in the convex combination
\begin{align}
x&=\sum_{i=0}^{k}t_{v_i}(x)\,v_i,
&
t_{v_i}(x)&\ge0,
&
\sum_{i=0}^{k}t_{v_i}(x)&=1.
\end{align}
Each weight lies in $[0,1]$, but the point is specified by the whole tuple of weights. The normalization leaves $k$ independent parameters, as appropriate for a $k$-dimensional simplex.

For each vertex $v$, we extend $t_v$ over the whole triangulation by setting it to zero on simplices that do not contain $v$. The definitions agree on shared faces, so $t_v$ is continuous on $\mathbb{CP}^N$. The region where $t_v>0$ is the open star of $v$, namely the union of the relative interiors of all simplices containing $v$.

On each ordered simplex, it is convenient to replace the barycentric weights by their cumulative sums, which we call Kuhn coordinates:
\begin{align}
y_j(x)&=\sum_{i=j}^{k}t_{v_i}(x),
\qquad j=1,\ldots,k.
\end{align}
These $k$ coordinates satisfy $1\ge y_1\ge\cdots\ge y_k\ge0$. This change of coordinates is invertible: introducing the auxiliary constants $y_0=1$ and $y_{k+1}=0$, we recover the barycentric weights as
\begin{align}
t_{v_i}(x)&=y_i(x)-y_{i+1}(x),
\qquad i=0,\ldots,k.
\end{align}

For a positive integer $m$, the \textit{edgewise subdivision} $K_m$ cuts every simplex of $K$ along the hyperplanes $y_i\in\frac1m\mathbb{Z}$ and $y_i-y_j\in\frac1m\mathbb{Z}$. Its vertices are the points at which every $m\,t_v$ is an integer. Its simplices are the Kuhn simplices
\begin{equation}\label{eq:sc-kuhn}
\Big\{x:\ m\,y(x)=b+z,\ 1\ge z_{\pi(1)}\ge\dots\ge z_{\pi(k)}\ge0\Big\}
\end{equation}
for $b\in\mathbb{Z}^k$ and permutations $\pi\in S_k$ that lie in the simplex, together with their faces. These subdivisions agree on common faces and form a simplicial complex that triangulates $\mathbb{CP}^N$~\cite{Freudenthal:1942,Edelsbrunner:2000egs}. $K_m$ is the $m$-fold edgewise subdivision of Ref.~\cite{Edelsbrunner:2000egs}, which Sec.~\ref{sec:lattices} applies to $\mathbb{CP}^3_{30}$ with $m=2$.

Let $T_k(K_m)$ be the set of $k$-simplices of $K_m$ and $f_d(K)$ the number of $d$-simplices of $K$. Each $d$-simplex of $K$ is cut into $m^d$ simplices, so $|T_k(K_m)|\le\binom{d+1}{k+1}f_d(K)\,m^d$. In particular, the number of qubits is $n=|T_2(K_m)|=O(m^d)$.

\begin{lemma}[Refinement and coordinates]\label{lem:sc-coords}
Let $q$ divide $m$ and set $a=m/q$. For a vertex $u$ of $K_q$, let $s_u$ be its barycentric coordinate function with respect to $K_q$, extended by zero.
\begin{enumerate}
\item[(a)] $K_m$ subdivides $K_q$: every simplex of $K_m$ lies in a simplex of $K_q$, and every simplex of $K_q$ is a union of simplices of $K_m$.
\item[(b)] If $x$ and $x'$ lie in a common simplex of $K_m$, then $|s_u(x)-s_u(x')|\le 2/a$ for every vertex $u$ of $K_q$.
\item[(c)] For every point $x$, some vertex $u$ of $K_q$ has $s_u(x)\ge 1/(d+1)$.
\end{enumerate}
\end{lemma}

Applying (b) with $(q,m)$ replaced by $(1,q)$ gives a bound for the coarse coordinates: if $x$ and $x'$ lie in a common simplex of $K_q$, then $|t_v(x)-t_v(x')|\le2/q$ for every vertex $v$ of $K$.

\begin{proof}
\textit{Idea.} $K_q$ and $K_m$ are cut out of each simplex of $K$ by the same families of hyperplanes, with spacings $1/q$ and $1/m$. A simplex of $K_m$ is therefore a small piece of a simplex of $K_q$, across which the coarse coordinates change by $O(q/m)$.

(a)~In the Kuhn coordinates of a simplex of $K$, the hyperplanes that cut out $K_q$ are among those that cut out $K_m$, since $\frac1q\mathbb{Z}\subset\frac1m\mathbb{Z}$.

(b)~By (a), $x$ and $x'$ lie in a common simplex $\tau$ of $K_q$, which is a face of a Kuhn simplex~\eqref{eq:sc-kuhn} with $q$ in place of $m$. On $\tau$ the barycentric coordinates with respect to $K_q$ are $1-z_{\pi(1)}$, $z_{\pi(j)}-z_{\pi(j+1)}$ and $z_{\pi(k)}$, where $z=q\,y-b$, and every other $s_u$ vanishes. Since $x$ and $x'$ lie in a common simplex of $K_m$, they lie in a common cube of side $1/m$ in the coordinates $y$. So each $z_i$ differs by at most $q/m$ between them, and each $s_u$ by at most $2q/m=2/a$.

(c)~The coordinates $s_u(x)$ are nonnegative, sum to one, and at most $d+1$ of them are nonzero.
\end{proof}

\begin{lemma}[Bounded geometry]\label{lem:sc-degree}
Let $N_K$ be the largest number of $d$-simplices of $K$ that contain a common vertex. Every vertex of $K_m$ lies in at most $\Gamma=2^d\,d!\,N_K$ $d$-simplices of $K_m$. Consequently:
\begin{itemize}
\item every vertex of $K_m$ has at most $D_K=d\,\Gamma$ neighbours;
\item every 3-simplex of $K_m$ shares a 4-simplex with at most $\Delta_K=4(d-3)\Gamma$ other 3-simplices.
\end{itemize}
None of these bounds depends on $m$.
\end{lemma}

\begin{proof}

Let $p$ be a vertex of $K_m$, and let $\sigma$ be the smallest simplex of $K$ containing $p$. Since $K_m$ subdivides $K$, every $d$-simplex of $K_m$ containing $p$ lies in a coarse $d$-simplex of $K$ containing $\sigma$. There are at most $N_K$ such coarse simplices: choosing any vertex $v$ of $\sigma$, all of them contain $v$, and $N_K$ bounds the number of coarse $d$-simplices incident to any vertex.

Now fix one such coarse simplex and use the rescaled Kuhn coordinates $z=m\,y$. The point $p$ has integer coordinates $z_p=m\,y(p)\in\mathbb Z^d$. Every fine simplex containing $p$ lies in a unit grid cube having $z_p$ as a corner. There are $2^d$ such cubes, since in each coordinate the cube extends either below or above $z_p$. Each cube is subdivided into $d!$ Kuhn simplices, one for each ordering of the $d$ local coordinates in Eq.~\eqref{eq:sc-kuhn}. Thus at most $2^d d!$ fine simplices containing $p$ lie in the chosen coarse simplex. Summing over the at most $N_K$ coarse simplices gives
\begin{align}
\#\{\text{$d$-simplices of $K_m$ containing $p$}\}
&\le N_K\,2^d d!
=\Gamma.
\end{align}

Each $d$-simplex that contains $p$ contributes at most $d$ neighbours of $p$, which gives $D_K$.

For $\Delta_K$: since $K_m$ triangulates a manifold, every 4-simplex is a face of a $d$-simplex. A 3-simplex $\rho$ lies in at most $\Gamma$ $d$-simplices. Each of these has $d-3$ four-dimensional faces that contain $\rho$, and each such 4-simplex has four 3-faces other than $\rho$.
\end{proof}

We note that repeated barycentric subdivision does not satisfy Lemma~\ref{lem:sc-degree}: each step at least doubles the degree of the vertices left from the previous step.

For a vertex $u$ of a triangulation $L$ of $\mathbb{CP}^N$ (we use $L=K$ and $L=K_q$), we write:
\begin{itemize}
\item $\mathrm{St}(u)$ for the closed star of $u$ and $\mathrm{Lk}(u)$ for its link;
\item $\mathrm{D}(u)$ for the subcomplex of simplices that do not contain $u$;
\item $\mathrm{St}(u)_m$ and $\mathrm{D}(u)_m$ for the subcomplexes of $K_m$ that subdivide $\mathrm{St}(u)$ and $\mathrm{D}(u)$, when $K_m$ subdivides $L$.
\end{itemize}
A simplex of $K_m$ lies in $\mathrm{St}(u)_m$ if and only if it lies in a simplex of $L$ that contains $u$. It lies outside $\mathrm{D}(u)_m$ if and only if the barycentric coordinate of $u$ with respect to $L$ is positive in its interior.

\begin{lemma}[Local acyclicity]\label{lem:sc-acyclic}
For every vertex $u$ of $L$, $H_k(\mathrm{St}(u)_m)=0$ for $k\ge1$, and $H^k(K_m,\mathrm{D}(u)_m)=0$ for $1\le k\le d-1$.
\end{lemma}

\begin{proof}
$\mathrm{St}(u)$ is a cone with apex $u$, hence contractible, and simplicial homology is invariant under subdivision. This gives the first claim.

For the second, subdivision invariance and simplicial excision, with $L=\mathrm{St}(u)\cup\mathrm{D}(u)$ and $\mathrm{St}(u)\cap\mathrm{D}(u)=\mathrm{Lk}(u)$, give $H^k(K_m,\mathrm{D}(u)_m)\cong H^k(L,\mathrm{D}(u))\cong H^k(\mathrm{St}(u),\mathrm{Lk}(u))$. Because $\mathrm{St}(u)$ is contractible, this group is $\tilde H^{k-1}(\mathrm{Lk}(u))$. Since $L$ triangulates a $d$-manifold, $\tilde H_{j}(\mathrm{Lk}(u))\cong H_{j+1}(\mathbb{CP}^N,\mathbb{CP}^N\setminus u)$ vanishes for $j\neq d-1$~\cite{Munkres:1984}. Over the field $\mathbb{Z}_2$, so does $\tilde H^{j}$.
\end{proof}

\subsection{The local-filling decoder}
\label{app:sc-decoder}

The decoder treats each small syndrome component separately. A small component is confined to the star of a vertex of $K_q$ (Lemma~\ref{lem:sc-local}), and the star is acyclic, so the component can be filled inside it. The decoded logical class of an error is the class of the error plus all these fillings. The key property is Lemma~\ref{lem:sc-bottleneck}: flipping a single qubit cannot change the decoded class unless a large component is present.

From now on, fix an integer $q>6(d+1)$, let $m\in q\mathbb{N}$, and set
\begin{equation}
a=\frac mq,\qquad r_0=\frac{a}{2(d+1)}=\frac{m}{2q(d+1)}~.
\end{equation}

\paragraph{Syndrome components.}
\begin{itemize}
\item \textit{$Z$-syndrome.} $\xi_Z=\partial E_Z$ is a 1-cycle. Its components are the edge sets of the connected components of the graph formed by its edges. Distinct components share no vertex, so each component is itself a 1-cycle.
\item \textit{$X$-syndrome.} $\xi_X=\delta E_X$ is a 3-cocycle. Call two of its 3-simplices adjacent when they are faces of a common 4-simplex; its components are the connected classes of this relation. Each component $c$ is again a cocycle: a 4-simplex with a face in $c$ has all of its faces in $\xi_X$ inside $c$, so $\delta c$ agrees with $\delta\xi_X=0$ there, and $\delta c$ vanishes on every other 4-simplex.
\item \textit{Small components.} A component is small if it has fewer than $r_0$ simplices. Let $B_Z$ be the set of errors $E$ whose $Z$-syndrome has a component that is not small, define $B_X$ in the same way, and set $B=B_X\cup B_Z$.
\end{itemize}

\begin{lemma}[Localization]\label{lem:sc-local}
Let $V$ be a set of vertices of $K_m$, and suppose that every vertex of $V$ can be reached from some $p\in V$ by an edge path of $K_m$ with fewer than $r_0$ edges. Let $u$ be a vertex of $K_q$ with $s_u(p)\ge1/(d+1)$. Then every simplex of $K_m$ whose vertices all lie in $V$ lies in a simplex of $K_q$ that contains $u$, and $s_u$ is positive in its interior.
\end{lemma}

\begin{proof}

By Lemma~\ref{lem:sc-coords}(b), $s_u$ changes by at most $2/a$ along each edge. Since every $x\in V$ can be joined to $p$ by a path of fewer than $r_0$ edges,
\begin{align}
s_u(x)&>\frac{1}{d+1}-\frac{2r_0}{a}=0.
\end{align}
Now let $\eta$ be a simplex of $K_m$ with vertices in $V$. Since $K_m$ refines $K_q$ by Lemma~\ref{lem:sc-coords}(a), $s_u$ is affine on $\eta$. Its values at all vertices are positive, so it is positive throughout $\eta$. Thus $\eta$ lies in the open star of $u$ in $K_q$, and in particular in a simplex of $K_q$ containing $u$.
\end{proof}

Both of small components of $X,Z$-syndromes satisfy the hypothesis of Lemma~\ref{lem:sc-local}, with $V$ their vertex set and $p$ any of their vertices:
\begin{itemize}
\item A $Z$-component $\gamma$ is connected, so a path inside $\gamma$ joins $p$ to any of its vertices, and such a path uses at most $|\gamma|<r_0$ edges.
\item For an $X$-component $c$: if two 3-simplices of $c$ are adjacent, every vertex of one is equal or joined by an edge to every vertex of the other. Following a chain of at most $|c|-1$ adjacent simplices, every vertex of $c$ is reached from $p$ by a path of at most $\max(1,|c|-1)<r_0$ edges.
\end{itemize}

\begin{definition}[Local-filling decoder]\label{def:sc-decoder}
Fix total orders on the vertices of $K_m$ and of $K_q$. For a small component, let $p$ be its first vertex and $u$ the first vertex of $K_q$ with $s_u(p)\ge1/(d+1)$, which exists by Lemma~\ref{lem:sc-coords}(c).
\begin{itemize}
\item By Lemma~\ref{lem:sc-local}, a small $Z$-component $\gamma$ is a 1-cycle of $\mathrm{St}(u)_m$, with $\mathrm{St}$ taken in $K_q$. By Lemma~\ref{lem:sc-acyclic} it bounds there; fix $F(\gamma)\in C_2(\mathrm{St}(u)_m)$ with $\partial F(\gamma)=\gamma$.
\item By Lemma~\ref{lem:sc-local}, a small $X$-component $c$ vanishes on $\mathrm{D}(u)_m$, so it is a relative 3-cocycle of $(K_m,\mathrm{D}(u)_m)$. Since $3\le d-1$, Lemma~\ref{lem:sc-acyclic} gives a 2-cochain $f(c)$ that vanishes on $\mathrm{D}(u)_m$ and has $\delta f(c)=c$; fix one.
\end{itemize}
For $E\notin B_Z$ and $E\notin B_X$ respectively, define the decoded classes
\begin{align}
\lambda_Z(E)&=\Big[E_Z+\sum_\gamma F(\gamma)\Big]\in H_2(K_m)\cong\mathbb{Z}_2,\\
\lambda_X(E)&=\Big[E_X+\sum_c f(c)\Big]\in H^2(K_m)\cong\mathbb{Z}_2,
\end{align}
with sums over the components of the respective syndromes. The decoder $\mathcal{D}_m$ measures all checks. If the syndrome is that of an error outside $B$, it applies $X\big(\sum_c f(c)\big)\,Z\big(\sum_\gamma F(\gamma)\big)$. Otherwise it applies any Pauli operator with the measured syndrome.
\end{definition}

After the correction, the residual operator is $X(E_X+\sum f)\,Z(E_Z+\sum F)$, a product of a 2-cocycle and a 2-cycle. On the code space it acts as $\bar X^{\lambda_X(E)}\bar Z^{\lambda_Z(E)}$, up to a phase. Hence, for $E\notin B$ and every operator $\bar\rho$ on the code space,
\begin{equation}\label{eq:sc-decode}
\mathcal{D}_m\big(P_E\bar\rho P_E^\dagger\big)=\bar X^{\lambda_X(E)}\bar Z^{\lambda_Z(E)}\,\bar\rho\,\big(\bar X^{\lambda_X(E)}\bar Z^{\lambda_Z(E)}\big)^\dagger~.
\end{equation}

The decoded class has a symmetry. Adding a 2-cycle $\ell_Z$ to $E_Z$, or a 2-cocycle $\ell_X$ to $E_X$, changes neither the syndromes nor the energy~\eqref{eq:sc-energy}, and it shifts $\lambda_Z$ by $[\ell_Z]$ and $\lambda_X$ by $[\ell_X]$.

\begin{lemma}[Bottleneck]\label{lem:sc-bottleneck}
Let $E$ and $E'$ differ by a single-qubit Pauli operator. If $E,E'\notin B_Z$, then $\lambda_Z(E)=\lambda_Z(E')$. If $E,E'\notin B_X$, then $\lambda_X(E)=\lambda_X(E')$.
\end{lemma}

\begin{proof}

\textit{$Z$-type.} If $E_Z=E'_Z$ there is nothing to prove. Otherwise $E'_Z=E_Z+e$ for a triangle $e$, and $\xi'_Z=\xi_Z+\partial e$.

Call a component of $\xi_Z$ affected if it contains a vertex of $e$. 
Let $A$ be the union of the affected components of $\xi_Z$, namely those that contain a vertex of $e$. The unaffected components share no vertices with either $A$ or $e$, so they remain unchanged in $\xi'_Z=\xi_Z+\partial e$ and retain their fillings. The remaining components of $\xi'_Z$ are the connected components of $A+\partial e$, where addition is modulo two.
Every component of $A+\partial e$ also contains a vertex of $e$. To see this, choose an edge $g$ in such a component. If $g\in\partial e$, the claim is immediate. Otherwise, $g$ belongs to an original affected component. Within that component, choose a path from an endpoint of $g$ to a vertex of $e$, stopping at the first such vertex encountered. Each edge of this path has at least one endpoint outside $e$, so none belongs to $\partial e$. Consequently, the entire path survives in $A+\partial e$, connecting $g$ to a vertex of $e$.

Let $\Gamma$ and $\Gamma'$ be the affected components of $\xi_Z$ and the new components of $\xi'_Z$. Then $\sum_{\Gamma'}\gamma'=\sum_\Gamma\gamma+\partial e$, so $\lambda_Z(E')-\lambda_Z(E)=[Z]$ with
\begin{equation}
Z=e+\sum_{\gamma\in\Gamma}F(\gamma)+\sum_{\gamma'\in\Gamma'}F(\gamma'),\qquad \partial Z=0~.
\end{equation}

Let $x_0$ be a vertex of $e$, and choose a vertex $w$ of $K$ with $t_w(x_0)\ge1/(d+1)$.
Let $\gamma\in\Gamma\cup\Gamma'$ have associated vertex $u\in K_q$, and choose a triangle $\eta$ of $K_m$ appearing in the filling $F(\gamma)$. Let $y\in\eta$. Both $\gamma$ and its chosen filling are supported in the subdivided closed star $\operatorname{St}(u)_m$, with the star taken in $K_q$. Choose a vertex $x_\gamma$ of $\gamma$ that is also a vertex of $e$, whose existence was established above. We then have:
\begin{itemize}
\item $y$ and $u$ lie in a common simplex of $K_q$, because $\eta$ lies in a coarse simplex containing $u$.
\item $x_\gamma$ and $u$ likewise lie in a common simplex of $K_q$, because $\gamma$ is supported in $\operatorname{St}(u)_m$.
\item $x_\gamma$ and $x_0$ are both vertices of the simplex $e$ of $K_m$.
\end{itemize}

Lemma~\ref{lem:sc-coords} then gives
\begin{equation}
|t_w(y)-t_w(x_0)|\le \frac2q+\frac2q+\frac2m\le\frac6q<\frac1{d+1}~,
\end{equation}
so $t_w(y)>0$. Points of $e$ satisfy the same bound. Every simplex of $Z$ therefore lies in a simplex of $K$ that contains $w$, so $Z$ is a 2-cycle of $\mathrm{St}(w)_m$, with $\mathrm{St}$ taken in $K$. By Lemma~\ref{lem:sc-acyclic}, $Z$ is a boundary, and $[Z]=0$.

\textit{$X$-type.} Now $E'_X=E_X+e$ and $\xi'_X=\xi_X+\mathcal{N}$, where $\mathcal{N}=\delta e$ is the set of 3-simplices that contain $e$. Call a 3-simplex near $e$ if it lies in $\mathcal{N}$ or is adjacent to a simplex of $\mathcal{N}$. The vertices of a near simplex $h$ and of $e$ lie in a common simplex of $K_m$: either $h$ contains $e$, or $h$ and a 3-simplex that contains $e$ are faces of a common 4-simplex, which then contains both $h$ and $e$.

Call a component of $\xi_X$ affected if it contains a near simplex. The unaffected components are adjacent neither to $\mathcal{N}$ nor to the affected ones, so they remain components of $\xi'_X$. The other components are those of $A+\mathcal{N}$, where $A$ is the union of the affected components. Each contains a near simplex. To see this, take $g\in A+\mathcal{N}$ with $g\notin\mathcal{N}$, and follow a chain of adjacent simplices inside its component in $A$ to the first near simplex $g^*$. If $g^*$ were in $\mathcal{N}$, its predecessor in the chain would be near, which contradicts the choice of $g^*$. So $g^*\notin\mathcal{N}$, and the whole chain lies in $A+\mathcal{N}$.

As before, $\lambda_X(E')-\lambda_X(E)=[Z']$ with $Z'=e+\sum_\Gamma f(c)+\sum_{\Gamma'}f(c')$ and $\delta Z'=0$. Take $x_0$ and $w$ as in the $Z$-type case. Let $y$ be a point of a simplex in the support of $f(c)$, for a component $c\in\Gamma\cup\Gamma'$ with associated vertex $u$ of $K_q$.
\begin{itemize}
\item The simplex lies outside $\mathrm{D}(u)_m$, so $y$ lies in a simplex of $K_q$ that contains $u$.
\item The component $c$ contains a near simplex $h$, as shown above. By Lemma~\ref{lem:sc-local}, applied to the vertex set of $c$, $h$ lies in a simplex of $K_q$ that contains $u$. Let $x_c$ be any vertex of $h$; it lies in that simplex.
\item $x_c$ and $x_0$ lie in a common simplex of $K_m$, by the property of near simplices above.
\end{itemize}
These are the three facts used in the $Z$-type case, with $x_c$ in place of $x_\gamma$, and Lemma~\ref{lem:sc-coords} gives $|t_w(y)-t_w(x_0)|\le6/q<1/(d+1)$, so $t_w(y)>0$. Points of $e$ satisfy the same bound. So $t_w$ is positive in the interior of every simplex of $Z'$, and $Z'$ vanishes on $\mathrm{D}(w)_m$, taken in $K$; it is a relative 2-cocycle. By Lemma~\ref{lem:sc-acyclic}, $Z'$ is the coboundary of a 1-cochain that vanishes on $\mathrm{D}(w)_m$, and $[Z']=0$.
\end{proof}

\begin{corollary}[Energy barrier]\label{cor:sc-barrier}
Every sequence of single-qubit Pauli operators that starts at the identity and ends at a nontrivial logical operator passes through an error that violates at least $r_0=m/(2q(d+1))$ checks. By Eq.~\eqref{eq:sc-energy}, the energy of that error is at least $2Jr_0=Jm/(q(d+1))$. The barrier is thus at least $r_0$ when measured in violated checks and at least $2Jr_0$ in energy.
\end{corollary}

\begin{proof}
The identity has $\lambda=(0,0)$. A nontrivial logical operator has no syndrome, and its $\lambda$ is its own nonzero class. By Lemma~\ref{lem:sc-bottleneck}, some intermediate error lies in $B$, so its syndrome has a component of at least $r_0$ violated checks.
\end{proof}

\subsection{Peierls bound}
\label{app:sc-peierls}

Let $\pi(E)\propto e^{-\beta\varepsilon(E)}$ be the Gibbs distribution on $C^2(K_m)\times C_2(K_m)$. Corollary~\ref{cor:sc-barrier} shows that a logical error must pass through a proliferation of non-small errors $B$; we now show that $B$ is exponentially rare.

\begin{lemma}[Peierls bound]\label{lem:sc-peierls}
If $e^{2\beta J}\ge2\max(D_K,\Delta_K^2)$, then $\pi(B_Z)\le2\,|T_1(K_m)|\,2^{-r_0}$ and $\pi(B_X)\le2\,|T_3(K_m)|\,2^{-r_0}$.
\end{lemma}

\begin{proof}

By Eq.~\eqref{eq:sc-energy}, $\pi$ is a product of a distribution of $E_X$ and a distribution of $E_Z$.

\textit{$Z$-type.} 
We first determine the distribution of the syndrome $\xi_Z=\partial E_Z$. For any 1-boundary $\xi$, choose one 2-chain with boundary $\xi$. Every other such filling is obtained by adding a 2-cycle, so each boundary has exactly $|Z_2(K_m)|$ fillings. This multiplicity is independent of $\xi$. Since the energy associated with the syndrome is $2J|\xi|$, its Gibbs distribution is therefore
\begin{align}
\Pr_\pi(\xi_Z=\xi)
&=\frac{1}{\mathcal Z_1}e^{-2\beta J|\xi|},
\end{align}
where $\mathcal Z_1$ is the normalization. The allowed syndromes are precisely the 1-cycles, because $H_1(\mathbb{CP}^N;\mathbb Z_2)=0$.

Now fix a connected 1-cycle $\gamma$, and let $\mathcal A_\gamma$ be the set of syndromes having $\gamma$ as a connected component. For $\xi\in\mathcal A_\gamma$, adding $\gamma$ simply removes that component:
\begin{align}
|\xi+\gamma|&=|\xi|-|\gamma|.
\end{align}
The resulting chain is still an allowed syndrome. Moreover, the removal map is injective, since adding $\gamma$ again recovers $\xi$. Consequently,
\begin{align}
\Pr_\pi(\xi_Z\in\mathcal A_\gamma)
&=
e^{-2\beta J|\gamma|}
\frac{\displaystyle\sum_{\xi\in\mathcal A_\gamma}
e^{-2\beta J|\xi+\gamma|}}{\mathcal Z_1}
\le e^{-2\beta J|\gamma|}.
\end{align}
The inequality follows because the numerator sums the weights of distinct allowed syndromes and is therefore at most $\mathcal Z_1$.

It remains to count the possible components. Fix an edge $g$ and one of its orientations. A connected 1-cycle with $\ell$ edges has even degree at every vertex, so it admits a closed walk that starts along $g$ in the chosen direction and traverses each edge exactly once. After the first edge, there are at most $D_K$ choices at each of the remaining $\ell-1$ steps. Thus at most $D_K^{\ell-1}$ connected 1-cycles of size $\ell$ contain $g$.

The event $B_Z$ occurs when some syndrome component has at least $r_0$ edges. Summing over its possible sizes and over an edge it contains, a union bound gives
\begin{align}
\pi(B_Z)
&\le |T_1(K_m)|
\sum_{\ell\ge r_0}D_K^{\ell-1}e^{-2\beta J\ell}\\
&\le |T_1(K_m)|
\sum_{\ell\ge r_0}\bigl(D_Ke^{-2\beta J}\bigr)^\ell\\
&\le 2\,|T_1(K_m)|\,2^{-r_0}.
\end{align}
In the last step we used the low-temperature condition $D_Ke^{-2\beta J}\le1/2$ and summed the geometric series.

\textit{$X$-type.} By the same reasoning, $\xi_X$ is distributed with weight $e^{-2\beta J|\xi|}$ on the 3-coboundaries. These are all the 3-cocycles because $H^3(\mathbb{CP}^N;\mathbb{Z}_2)=0$, and a component $c$ occurs with probability at most $e^{-2\beta J|c|}$. A connected set of $\ell$ 3-simplices that contains a given one has a spanning tree. The closed walk around this tree has $2(\ell-1)$ steps in a graph of degree at most $\Delta_K$, and it determines the set, so there are at most $\Delta_K^{2(\ell-1)}$ such sets. Hence $\pi(B_X)\le|T_3(K_m)|\sum_{\ell\ge r_0}(\Delta_K^2e^{-2\beta J})^\ell\le2\,|T_3(K_m)|\,2^{-r_0}$.
\end{proof}

\subsection{Thermal dynamics}
\label{app:sc-davies}

We now turn to the dynamics. The quantum memory reduces to a classical Markov chain on Pauli errors, to which the bounds above apply.

\begin{definition}[Thermal bath]\label{def:sc-bath}
The code Hamiltonian $H=-J\sum_e{\cal S}^X_e-J\sum_{s_3}{\cal S}^Z_{s_3}$ ($J>0$) on $K_m$ is weakly coupled to independent baths at inverse temperature $\beta$, through every single-qubit Pauli operator $\sigma^\alpha_f$ ($f\in T_2(K_m)$, $\alpha\in\{x,y,z\}$). In the weak-coupling (Davies) limit~\cite{Davies:1974} the state obeys $\dot\rho=\mathcal{L}(\rho)$ with
\begin{equation}
\begin{split}
\mathcal{L}(\rho)={}&-i[H+H_{\rm LS},\rho]+\sum_{f,\alpha,\omega}h_\alpha(\omega)\Big(A_{f\alpha}(\omega)\rho A_{f\alpha}(\omega)^\dagger\\
&-\tfrac12\big\{A_{f\alpha}(\omega)^\dagger A_{f\alpha}(\omega),\rho\big\}\Big)~.
\end{split}
\end{equation}

Here $f\in T_2(K_m)$ labels a triangle, and hence a physical qubit, while $\alpha\in\{x,y,z\}$ specifies the Pauli operator acting on that qubit: $\sigma_f^x=X_f$, $\sigma_f^y=Y_f$, and $\sigma_f^z=Z_f$. We use units with $\hbar=1$.

To define the jump operators, write the spectral decomposition of the code Hamiltonian as
\begin{align}
H&=\sum_\epsilon \epsilon\,\Pi_\epsilon,
\end{align}
where $\Pi_\epsilon$ projects onto the eigenspace of energy $\epsilon$. The index $\omega$ runs over the Bohr frequencies, namely energy differences between eigenspaces of $H$, including $\omega=0$. We decompose each single-qubit Pauli operator into components with a definite energy change:
\begin{align}
A_{f\alpha}(\omega)
&=\sum_{\epsilon-\epsilon'=\omega}
\Pi_{\epsilon'}\sigma_f^\alpha\Pi_\epsilon.
\end{align}
Thus $A_{f\alpha}(\omega)$ maps energy $\epsilon$ to energy $\epsilon-\omega$: positive $\omega$ corresponds to energy released by the system, and negative $\omega$ to energy absorbed from the bath.

The nonnegative coefficient $h_\alpha(\omega)$ sets the rate of the corresponding dissipative transition. These rates are determined by the bath correlation functions and satisfy the thermal detailed-balance (or KMS) condition
\begin{align}
h_\alpha(-\omega)&=e^{-\beta\omega}h_\alpha(\omega).
\end{align}
For $\omega>0$, this means that an energy-increasing transition is suppressed relative to its reverse by the Boltzmann factor $e^{-\beta\omega}$.

The bath also produces a coherent correction to the Hamiltonian, called the Lamb shift:
\begin{align}
H_{\rm LS}
&=\sum_{f,\alpha,\omega}
S_\alpha(\omega)\,
A_{f\alpha}(\omega)^\dagger A_{f\alpha}(\omega).
\end{align}
Here $S_\alpha(\omega)\in\mathbb R$ are energy-shift coefficients determined by the bath correlation functions. Unlike $h_\alpha(\omega)$, they describe coherent energy shifts rather than dissipative transition rates. Our notation takes both sets of coefficients to be independent of the qubit location $f$.

Finally, we define
\begin{align}
h_{\max}&=\max_{\alpha,\omega}h_\alpha(\omega),
\end{align}
where the maximum is over the three Pauli channels and the finitely many relevant Bohr frequencies. This is the thermal-bath setting of Refs.~\cite{doi:10.1142/S1230161210000023,Temme:2017}.

\end{definition}

\begin{lemma}[Classical reduction]\label{lem:sc-classical}
Let $\bar\rho$ be an operator on the code space and $p_0$ a probability distribution on errors $E$. Then
\begin{equation}
e^{t\mathcal{L}}\Big(\sum_Ep_0(E)\,P_E\bar\rho P_E^\dagger\Big)=\sum_Ep_t(E)\,P_E\bar\rho P_E^\dagger~,
\end{equation}
where $p_t$ is the law at time $t$ of a continuous-time Markov chain started from $p_0$. In this chain, $E$ jumps to $E\oplus(f,\alpha)$ at rate $h_\alpha\big(\varepsilon(E)-\varepsilon(E\oplus(f,\alpha))\big)$, where $(f,x)$ flips $E_X$ on $f$, $(f,z)$ flips $E_Z$ on $f$, and $(f,y)$ flips both. The chain is reversible with respect to $\pi$.
\end{lemma}

\begin{proof}

$P_E\bar\rho P_E^\dagger$ lies in the common eigenspace of the checks with the syndrome of $E$, on which $H$, $H_{\rm LS}$ and $A_{f\alpha}(\omega)^\dagger A_{f\alpha}(\omega)$ act as scalars. For the last two, $\sigma^\alpha_f\Pi_{\epsilon'}\sigma^\alpha_f$ is a sum of syndrome projectors. Moreover, $A_{f\alpha}(\omega)P_E\bar\rho P_E^\dagger A_{f\alpha}(\omega)^\dagger$ equals $\sigma^\alpha_fP_E\bar\rho P_E^\dagger\sigma^\alpha_f$ if $\omega=\varepsilon(E)-\varepsilon(E\oplus(f,\alpha))$, and vanishes otherwise. Hence
\begin{equation*}
\begin{split}
\mathcal{L}(P_E\bar\rho P_E^\dagger)=\sum_{f,\alpha}&h_\alpha\big(\varepsilon(E)-\varepsilon(E\oplus(f,\alpha))\big)\\
&\times\big(P_{E\oplus(f,\alpha)}\bar\rho P_{E\oplus(f,\alpha)}^\dagger-P_E\bar\rho P_E^\dagger\big),
\end{split}
\end{equation*}
up to phases that cancel. The right-hand side of the claim therefore solves $\dot\rho=\mathcal{L}(\rho)$ with the correct initial value. Reversibility follows from the KMS condition.
\end{proof}

\subsection{Proof of Theorem~\ref{thm:selfcorrection}}
\label{app:sc-proof}

\begin{definition}[Thermal encoding]\label{def:sc-encoding}
Let
\begin{align}
G&=\{E\notin B:\lambda_X(E)=\lambda_Z(E)=0\}
\end{align}
be the set of errors outside $B$ whose correction by $\mathcal D_m$ leaves no logical error. We sample errors from the Gibbs distribution $\pi$, conditioned on belonging to $G$. Explicitly,
\begin{align}
\pi(E\mid G)
&=\frac{\pi(E)}{\pi(G)}
=\frac{e^{-\beta\epsilon(E)}}
{\displaystyle\sum_{F\in G}e^{-\beta\epsilon(F)}},
\qquad E\in G,
\end{align}
where $\pi(G)=\sum_{F\in G}\pi(F)$ and $\epsilon(E)$ is the energy cost of the error $P_E$. The thermal encoding of a logical state $\bar\rho$ is
\begin{align}
\rho_\beta(\bar\rho)
&=\sum_{E\in G}\pi(E\mid G)\,P_E\bar\rho P_E^\dagger.
\end{align}
\end{definition}

Thus, $\rho_\beta(\bar\rho)$ is obtained by applying a thermally distributed, correctable Pauli error to the encoded state $\bar\rho$. It can contain thermal excitations, but decoding returns the original logical state:
\begin{align}
\mathcal D_m\bigl(\rho_\beta(\bar\rho)\bigr)&=\bar\rho.
\end{align}
The choice of decoder enters through the set $G$.

For comparison, let $\bar\rho_{\rm mix}$ be the maximally mixed state on the code space. If we remove the conditioning and average over all Pauli errors, we recover the full Gibbs state:
\begin{align}
\sum_E\pi(E)\,P_E\bar\rho_{\rm mix}P_E^\dagger
&=\frac{e^{-\beta H}}{\operatorname{tr}e^{-\beta H}}.
\end{align}

\begin{proof}[Proof of Theorem~\ref{thm:selfcorrection}]

Take $\beta_0=(2J)^{-1}\ln\big(2\max(D_K,\Delta_K^2)\big)$, which by Lemma~\ref{lem:sc-degree} does not depend on $m$, and let $\beta\ge\beta_0$.

\textit{Step 1: reduction to a classical failure probability.} By Lemma~\ref{lem:sc-classical}, $e^{t\mathcal{L}}\rho_\beta(\bar\rho)=\sum_Ep_t(E)P_E\bar\rho P_E^\dagger$, where $p_t$ is the law of the chain started from $\pi(\cdot\,|\,G)$. By Eq.~\eqref{eq:sc-decode},
\begin{equation}
\big\|\mathcal{D}_m\big(e^{t\mathcal{L}}\rho_\beta(\bar\rho)\big)-\bar\rho\big\|_1\le2\,\eta(t)~,
\end{equation}
where $\eta(t)=\Pr\big[E_t\in B\ \text{or}\ \lambda(E_t)\neq(0,0)\big]$ is the failure probability.

\textit{Step 2: comparison with equilibrium dynamics.}
Consider the same Markov dynamics as in Step 1, but now draw the initial error $E_0$ from the full Gibbs distribution $\pi$. Denote this process by $(E_s)_{s\ge0}$ and its classical generator by $Q$. For $E\ne E'$, the entry $Q(E,E')$ is the transition rate from $E$ to $E'$, while the diagonal entries make each row sum to zero. Since $\pi$ is an equilibrium distribution, $E_s\sim\pi$ for every $s\ge0$. We write $\Pr_\pi$ for probabilities in this stationary process.

Define
\begin{align}
A_Z={}&\{E_t\in B_Z\}\nonumber\\
&\cup\{E_0,E_t\notin B_Z,\ 
\lambda_Z(E_t)\ne\lambda_Z(E_0)\},
\end{align}
and define $A_X$ analogously. Thus $A_Z$ occurs if the final $Z$ syndrome has a large component, or if both endpoint syndromes have only small components but their decoded $Z$ classes differ.

Conditioning this equilibrium process on $E_0\in G$ gives exactly the initial distribution $\pi(\cdot\mid G)$ used in Step 1. Under this condition, both initial logical classes are zero, so a decoding failure at time $t$ implies $A_X\cup A_Z$. Therefore,
\begin{align}
\eta(t)
&\le \Pr_\pi(A_X\cup A_Z\mid E_0\in G)\nonumber\\
&\le
\frac{\Pr_\pi(A_X)+\Pr_\pi(A_Z)}{\pi(G)}.
\end{align}

To bound the denominator, recall that multiplying an error by a logical Pauli preserves its syndrome and energy but shifts its decoded logical class. The four classes outside $B$ therefore have equal Gibbs probability, giving
\begin{align}
\pi(G)&=\frac{1-\pi(B)}{4}.
\end{align}
If $\pi(B)\le1/2$, then $\pi(G)\ge1/8$, and hence
\begin{align}
\eta(t)
&\le8\bigl(\Pr_\pi(A_X)+\Pr_\pi(A_Z)\bigr).
\end{align}
The same bound holds trivially when $\pi(B)>1/2$: stationarity gives $\Pr_\pi(A_X)+\Pr_\pi(A_Z)\ge\pi(B)>1/2$, so the right-hand side exceeds one, whereas $\eta(t)\le1$.

\textit{Step 3: bounding $\Pr_\pi(A_Z)$.}
Define $\phi(E)=(-1)^{\lambda_Z(E)}$ outside $B_Z$ and $\phi(E)=0$ on $B_Z$. Thus $\phi$ records the decoded $Z$ class by the signs $+1$ and $-1$, with zero marking configurations in $B_Z$.

The Dirichlet form
\begin{align}
\mathcal E(\phi)
&=\frac12\sum_E\sum_{E'\ne E}
\pi(E)Q(E,E')
\bigl(\phi(E')-\phi(E)\bigr)^2
\end{align}
measures how strongly the allowed transitions change $\phi$, averaged over equilibrium configurations.

Every allowed jump is a single-qubit Pauli change. If both endpoints lie outside $B_Z$, Lemma~\ref{lem:sc-bottleneck} says that their decoded $Z$ classes agree, so $\phi$ does not change. If both endpoints lie inside $B_Z$, their $\phi$ values are both zero. Hence only transitions across the boundary of $B_Z$ contribute, and each has squared change equal to one. Detailed balance,
$\pi(E)Q(E,E')=\pi(E')Q(E',E)$, makes the incoming and outgoing contributions equal. Therefore,
\begin{align}
\mathcal E(\phi)
&=\sum_{\substack{E\in B_Z\\E'\notin B_Z}}
\pi(E)Q(E,E')\nonumber\\
&\le\sum_{E\in B_Z}\pi(E)
\sum_{E'\ne E}Q(E,E')
\le R\,\pi(B_Z).
\end{align}
Here $R=3n h_{\max}$ bounds the total outgoing rate: there are at most three Pauli jumps per qubit, each with rate at most $h_{\max}$.

For a stationary reversible chain, the mean-squared change over time $t$ satisfies
\begin{align}
\mathbb E_\pi\!\left[
\bigl(\phi(E_t)-\phi(E_0)\bigr)^2
\right]
&\le2t\,\mathcal E(\phi)
\le2tR\,\pi(B_Z).
\end{align}
This follows from the spectral decomposition of $Q$, applying $1-e^{-\mu t}\le\mu t$ to each mode with decay rate $\mu\ge0$~\cite{LevinPeres:2017}.

Let $D_Z$ denote the event that $E_0$ and $E_t$ both lie outside $B_Z$ but have different decoded $Z$ classes. On this event, $\phi$ changes from $+1$ to $-1$ or vice versa, so its squared change is four. Consequently,
\begin{align}
\Pr_\pi(D_Z)
&\le\frac14\,
\mathbb E_\pi\!\left[
\bigl(\phi(E_t)-\phi(E_0)\bigr)^2
\right]
\le\frac{tR}{2}\,\pi(B_Z).
\end{align}
Finally, $A_Z$ is the disjoint union of $D_Z$ and $\{E_t\in B_Z\}$. Since stationarity gives $\Pr_\pi(E_t\in B_Z)=\pi(B_Z)$,
\begin{align}
\Pr_\pi(A_Z)
&\le\pi(B_Z)\bigl(1+tR/2\bigr).
\end{align}
The same argument applies to $A_X$.

\textit{Step 4: conclusion.} Combining Steps 2 and 3 with Lemma~\ref{lem:sc-peierls},
\begin{equation}
\eta(t)\le16\big(|T_1(K_m)|+|T_3(K_m)|\big)\big(1+\tfrac32\,n\,h_{\max}t\big)\,2^{-r_0}~.
\end{equation}
Since $n$ and $|T_k(K_m)|$ are $O(m^d)$ with constants that depend only on $K$, this is the bound of Theorem~\ref{thm:selfcorrection}.
\end{proof}

\section{How the logical error rates are obtained}
\label{app:sim}

This appendix describes step by step how the logical error rates of
Secs.~\ref{sec:simulation} and~\ref{sec:cczprep} are computed.

\subsection{Common pipeline}
\label{app:pipeline}

\paragraph{Circuit, detectors and observable.} Each experiment is a
Stim circuit~\cite{Gidney:2021stim} built directly from the list of
simplices: qubit initializations, layers of CNOT gates scheduled by the
greedy edge colouring of Sec.~\ref{sec:circuits}, and noisy single-qubit
measurements. On top of
the circuit we declare \textit{detectors} and one \textit{logical
observable}. 
We use detectors to test consistency between measurement outcomes. Each detector compares a measured check value with either its known initial value or the value reconstructed from the final data-qubit readout. The detector is $0$ when the values agree and $1$ when they differ. We also read out the logical operator by taking the parity of the final data measurements over its support; this parity defines the logical observable. In these memory experiments, its expected value is known from the initial state. For each circuit, we verify that all detectors and the logical observable are $0$ in every noiseless shot, so any nonzero value signals a fault.

\paragraph{Detector error model.} Stim propagates every elementary
fault of the noise model through the circuit and records which
detectors it flips and whether it flips the observable. The result is the detector error model: a list
of independent mechanisms $j$ with probabilities $p_j$, written as a
binary matrix $H$ (detectors $\times$ mechanisms) and a binary row
vector $o$ whose entry $o_j$ says whether mechanism $j$ flips the
observable. 
We construct the binary matrix $H$ from Stim's detector error model, retaining the correlations between detectors flipped by the same error mechanism (\texttt{decompose\_errors=False}). This conversion is exact for the depolarizing and bit- or phase-flip channels used in our circuits. 
$H$ exactly represents the assumed circuit noise model, which applies independent Pauli noise channels after each operation.

\paragraph{Sampling and scoring.} For each shot Stim samples the
detector outcomes $s$ and the actual flip $o_{\rm act}$ of the
observable. The decoder sees only $s$ and returns a set of mechanisms
$\hat e$ with $H\hat e=s$; its predicted flip is $o\cdot\hat e$ (mod 2).
The shot is a logical error when the prediction differs from
$o_{\rm act}$, i.e.\ when the actual fault set and the correction
together act as a nontrivial logical operator. The logical error rate is
the fraction of failed shots, $p_L=N_{\rm fail}/N$.

\paragraph{Decoders.} We use min-sum belief propagation (BP) with
scaling factor $0.9$ and a serial schedule from the \texttt{ldpc}
package~\cite{Roffe:2020bposd}, run on $H$ with priors $p_j$. Where
stated, BP is followed by ordered-statistics decoding (OSD) of the
combination-sweep type and order~7, which returns a solution of
$H\hat e=s$ even when BP does not converge.

\subsection{One-round memory benchmark}
\label{app:memory}

\paragraph{Circuit and noise.} One syndrome round of the
$[[783,1,4]]$ code with one bare ancilla per check, as in
Sec.~\ref{sec:circuits}: a $Z$ sub-round of $15$ CNOT layers (data
$\to$ ancilla) for the $2\,110$ $Z$-checks, then an $X$ sub-round of
$17$ layers (ancilla $\to$ data, with Hadamard gates on the $153$
$X$-ancillas), a measurement of all $2\,263$ ancillas, and a
readout of all $783$ data qubits. Every CNOT is followed by two-qubit
depolarizing noise of strength $p$, and every measurement, including
the final readout, is preceded by a flip of probability $p$.
Initialization and idling are noiseless.

\paragraph{$Z$ basis (Table~\ref{tab:sim783}).} The data qubits start
in $|0\rangle$ and are read out in the $Z$ basis. Every $Z$-check then
has a known first outcome, giving $2\,110$ detectors, and a further
$2\,110$ detectors compare each $Z$-check outcome with the parity of the
readout on its four triangles. The $X$-check outcomes are random in this
round and are not detectors. The observable is the readout parity on the
four triangles of an empty tetrahedron, a weight-4 $\bar Z$.
The detector error model has $4\,220$ detectors and $13\,223$ mechanisms, and is decoded by BP with $100$ iterations. 

Measuring the full, redundant set of $Z$-checks helps the decoder distinguish data errors from faulty check readouts. Ideal syndromes obey parity relations—the metachecks of Sec.~\ref{sec:singleshot}—that measurement errors can violate. 
These relationships are already encoded in the circuit's detector error model, so the decoder uses the redundancy without imposing metachecks as separate constraints.
Only errors with an $X$ component can flip $\bar Z$, and the $Z$-checks detect this component. After decoding, an undetected residual $X$ error that flips $\bar Z$ must be a nontrivial $X$-type logical operator, whose minimum weight is $d_X\ge91$.

\paragraph{$X$ basis (Table~\ref{tab:simX}).} The data qubits start in
$|+\rangle$ and are read out in the $X$ basis, with a phase flip of
probability $p$ before the readout. The detectors are the $153$
first-round $X$-check outcomes and their $153$ comparisons with the
readout.
We read out $\bar X$ using a weight-$149$ logical $X$ operator that anticommutes with the weight-$4$ logical $\bar Z$ defined above. The observable is the parity of the final $X$-basis measurements on the $149$ data qubits supporting this operator.
The detector error model has $306$ detectors and $5\,395$
mechanisms and is decoded by BP+OSD. Only $Z$-type faults flip $\bar X$,
and they face the weight-4 logicals. Each triangle lies in three
$X$-checks but, on average, in $10.8$ $Z$-checks, so most of these
$Z$-type faults come from the CNOTs of the $Z$ sub-round, through
depolarizing noise on the data qubits and through phase errors on
$Z$-ancillas that spread back to the data.

\subsection{$|CCZ\rangle$ preparation}
\label{app:ccz}

\paragraph{Circuit and noise of the proxy.} The proxy simulates one of
the three blocks of the magic-state preparation in the $X$ basis, where
only $Z$-type faults matter. All data qubits and the $153$
$X$-ancillas start in $|+\rangle$. 
Each data qubit receives single-qubit depolarizing noise of strength $p$ once for preparation and once for each physical $CCZ$ gate acting on it. Thus, the qubit on triangle $t$ in block~1 receives $1+\ell_t$ noise applications, where $\ell_t$ counts the CCZ gates acting on that qubit, as defined in Sec.~\ref{sec:cczprep}.
One
$X$-check round of $17$ CNOT layers with two-qubit depolarizing noise
$p$ follows, and the ancillas and data are read out in the $X$ basis,
each after a phase flip of probability $p$. Detectors are the $153$
first-round $X$-checks and their comparisons with the readout, and the
observable is the readout parity over an $\bar X$ of $149$ triangles.
The model has $306$ detectors and $3\,285$ mechanisms. 

\paragraph{With the $Z$-check round.} Step~1 of the protocol of
Sec.~\ref{sec:universal} prepares $|\bar+\rangle$ by measuring the
$Z$-checks, which the proxy above omits. We add this measurement between
the preparation and the $CCZ$ legs, with one bare ancilla per check and
the noise of Sec.~\ref{app:memory}. 
We simulate two alternative circuits for the $Z$-check preparation step. The first measures all $2\,110$ checks using the $Z$ sub-round of Sec.~\ref{app:memory}, with $8\,440$ CNOTs in $15$ layers. The second measures only an independent set of $646$ checks, with $2\,584$ CNOTs in six layers.

\paragraph{Residual $X$ errors and kickback.} 
To estimate the $X$ errors left by each preparation circuit, we simulate it separately in the $Z$ basis, where the check outcomes define detectors, and decode with BP+OSD. Comparing the actual $X$ error with the decoder's predicted correction gives the residual error on each triangle. Residual $X$ operators that commute with all $Z$-checks act trivially on $|\bar+\rangle$ and are omitted. 

We then estimate how these residual errors propagate through the $CCZ$ gates to block~1. For each 6-simplex, an $X$ error on its middle or back triangle produces an additional $CZ$ error on the other two qubits. After Pauli twirling, this gives a $Z$ error on the front triangle with probability $\frac12$. We sum these contributions over the 6-simplices and apply the resulting rates as independent $Z$ noise on block~1, before the depolarizing noise representing the $CCZ$ gates. This approximation neglects correlations between blocks and, like the rest of the proxy, idle noise.

\paragraph{Two-class decoding.}
We compare the two possible logical outcomes by appending the observable row $o$ to $H$ and decoding each syndrome twice with BP+OSD, using $60$ BP iterations per decode. Setting the additional syndrome bit to $c=0$ or $1$ gives a candidate error pattern $\hat e_c$ satisfying $H\hat e_c=s$ and $o\cdot\hat e_c=c$. We compare the candidates using their log-likelihood weights,
\begin{align}
W_c=\sum_{j\in\hat e_c}\log\frac{1-p_j}{p_j},
\end{align}
where $p_j$ is the probability of mechanism $j$. The decoder predicts the logical flip corresponding to the smaller weight.
We use the weight gap $|W_0-W_1|$ as a confidence measure and reject shots whose gap is below a threshold $g$. The rejection rate is the fraction of all shots rejected, while the accepted error rate is the fraction of accepted shots with an incorrect logical prediction.

\bibliographystyle{utphys}
\bibliography{biblio.bib,mybib_merge}
\end{document}